\documentclass[prd,twocolumn,superscriptaddress,nofootinbib,eqsecnum,showpacs]{revtex4}

\usepackage{amsmath}
\usepackage{amsfonts}
\usepackage{amssymb}
\usepackage{bm}
\usepackage{color}
\usepackage[colorlinks]{hyperref}

\definecolor{CiteColor}{rgb}{0,0.5,0}
\hypersetup{citecolor=CiteColor}
\definecolor{RefColor}{rgb}{0.55,0,0}
\hypersetup{linkcolor=RefColor}

\newcommand{\ud}{\mathrm{d}}
\newcommand{\uD}{\mathrm{D}}
\newcommand{\beq}{\begin{equation}}
\newcommand{\eeq}{\end{equation}}
\newcommand{\bseq}{\begin{subequations}}
\newcommand{\eseq}{\end{subequations}}

\newcommand{\IAP}{\affiliation{Institut d'Astrophysique de Paris,
   UMR 7095 du CNRS,\\ Universit\'e Pierre \& Marie Curie, 98$^{\text{bis}}$
   Bvd. Arago, 75014 Paris, France}}
\newcommand{\Maryland}{\affiliation{Maryland Center for Fundamental
    Physics \& Joint Space-Science Institute,\\ Department of Physics,
    University of Maryland, College Park, MD 20742, USA}}

\begin{document}

\title{First Law of Mechanics for Black Hole Binaries with Spins}

\author{Luc Blanchet} \IAP
\author{Alessandra Buonanno} \Maryland 
\author{Alexandre Le Tiec} \Maryland

\date{\today}

\begin{abstract}
We use the canonical Hamiltonian formalism to generalize to spinning point particles the first law of mechanics established for binary systems of non-spinning point masses moving on circular orbits [Le Tiec, Blanchet, and Whiting, Phys. Rev. D \textbf{85}, 064039 (2012)]. We find that the redshift observable of each particle is related in a very simple manner to the canonical Hamiltonian and, more generally, to a class of Fokker-type Hamiltonians. Our results are valid through linear order in the spin of each particle, but hold also for quadratic couplings between the spins of different particles. The knowledge of spin effects in the Hamiltonian allows us to compute spin-orbit terms in the redshift variable through 2.5PN order, for circular orbits and spins aligned or anti-aligned with the orbital angular momentum. To describe extended bodies such as black holes, we supplement the first law for spinning point-particle binaries with some ``constitutive relations'' that can be used for diagnosis of spin measurements in quasi-equilibrium initial data.
\end{abstract}


\maketitle

\section{Introduction}
\label{sec:intro}

The need to develop faithful template waveforms for the search of gravitational waves  from compact-object binary systems (compact binaries, for short) with current interferometric detectors \cite{LIGO,Virgo} has led, during the last years, to synergies and unexpected common ground among previously distinct research areas. The two main analytical frameworks used to study the relativistic dynamics of compact binaries are (i) the post-Newtonian (PN) approximation \cite{Bl.06,SaTa.03}, which describes the inspiraling motion beyond the Newtonian limit, in the small velocity and weak-field regime ($v \ll c$), and (ii) black hole perturbation theory \cite{Po.al.11}, which describes both the weak-field and strong-field dynamics in the small mass-ratio limit ($m_1 \ll m_2$). The other main research area in the study of the two-body problem is numerical relativity \cite{Ce.al.10}, which aims at describing the strong-field regime by solving numerically the exact Einstein field equations. Furthermore, in order to build templates for the search of gravitational waves, the effective-one-body (EOB) approach \cite{BuDa.99,BuDa.00} combines the information from those different techniques in a flexible and effective way, and can provide accurate merger templates for advanced LIGO and Virgo. As said above, several studies \cite{Aj.al.08,DaNa.09,De.08,Bl.al.10,Bl.al2.10,Da.10,Ba.al.10,Le.al.11,Pa.al.11,Da.al.12,Le.al2.12,Ba.al.12,Ak.al.12} at the interface between those different frameworks have improved our knowledge of the two-body dynamics and gravitational-wave emission.
 
Recently, Le Tiec, Blanchet, and Whiting \cite{Le.al.12} (henceforth, Paper I) derived a ``First Law of Mechanics'' for binary systems of point particles moving on exact circular orbits (compatible with an helical Killing symmetry). Using the first law, the authors found a very simple relation between the PN binding energy of the binary and Detweiler's redshift observable \cite{De.08} --- which can also be interpreted as the particle's Killing energy associated with the helical symmetry. This relation allowed the computation of previously unknown high-order PN terms in the circular-orbit binding energy by using the redshift observable computed numerically within the gravitational self-force (GSF) formalism \cite{De.08,Sa.al.08,Bl.al.10,Bl.al2.10}. Moreover, the results in Paper I led to the following applications: 
\begin{enumerate}
	\item Reference \cite{Le.al2.12} computed the \textit{exact} binding energy and angular momentum of two non-spinning compact objects (modeled as point particles) moving on circular orbits, at leading order beyond the test-particle approximation, and recovered the exact frequency shift of the Schwarzschild innermost stable circular orbit induced by the conservative piece of the GSF \cite{BaSa.09}; 
	\item Reference \cite{Ba.al.12} built upon the works \cite{Le.al.12,Le.al2.12,Da.10,Ba.al.10} and derived the exact expressions of the EOB metric components through first order in the (symmetric) mass ratio. Quite interestingly, the results in Refs.~\cite{Le.al.11,Da.al.12,Le.al2.12} strongly suggest that the domain of validity of black hole perturbation theory calculations may extend well beyond the extreme mass-ratio limit.
\end{enumerate}
 
Given the relevance of the first law of binary mechanics in enhancing our knowledge of the two-body dynamics in the comparable-mass case by using information from the perturbative GSF framework, we generalize it here to point particles carrying spins. We derive the first law of mechanics using the canonical Arnowitt-Deser-Misner (ADM) formalism \cite{Ar.al.62} applied to spinning point particles \cite{St.al2.08,StSc.09}. In the PN context, both the ADM formalism and the harmonic-coordinate approach have been developed up to high orders in the approximation, in order to compute the conservative part of the orbital dynamics of compact binaries. For non-spinning point masses, the Hamiltonian \cite{JaSc.98,Da.al.00,Da.al.01} and the harmonic-coordinates equations of motion \cite{BlFa2.00,BlFa.01,deA.al.01,BlIy.03,Bl.al.04} are known through 3PN order. More recently, partial results at 4PN order have been reported \cite{FoSt.12,JaSc.12}. The logarithmic contributions to the conservative dynamics at 4PN and 5PN orders are also known \cite{Bl.al2.10,Le.al.12}. High-order spin-orbit effects have been computed in harmonic coordinates \cite{Ta.al.01,Fa.al.06,Ma.al.12}, in the Hamiltonian \cite{Da.al.08,St.al2.08,HaSt.11}, and within the Effective Field Theory (EFT) approach \cite{Le2.10}. See Ref.~\cite{St.11} for a review of the canonical ADM formalism for spinning point particles, and its application in PN theory. In this paper we shall not consider spin-spin couplings or higher-order effects in the spins. High-order spin-spin effects have been computed in the Hamiltonian \cite{St.al.08,St.al2.08,St.al3.08,He.al.10,HaSt2.11} and using EFT techniques \cite{Po.06,PoRo.06,PoRo.08,PoRo2.08,Le.10,Le2.12}.

The paper is organized as follows. In Secs.~\ref{sec:hamilton} and \ref{sec:Fokker} we review the Lagrangian and Hamiltonian formalisms for systems of point particles carrying spins. In particular, we derive a crucial relationship between the redshift observable and the variation of a general (Fokker-type) Hamiltonian, valid at linear order in the spins. In Sec.~\ref{sec:proof} we derive the first law of mechanics for binary systems of spinning point particles using the canonical ADM formalism, and discuss some mathematical and physical consequences of this law. Then, in Sec.~\ref{sec:PN}, we employ the first law of mechanics and the ADM Hamiltonian to compute the spin-orbit contributions to the redshift observable through the (leading plus sub-leading) 2.5PN order. We find full agreement between our results and those recently obtained from a direct computation based on the near-zone PN metric \cite{Fr.al.12}. In Sec.~\ref{sec:corot} we discuss the first law in the particular case of binary black holes in corotation. Finally, Sec.~\ref{sec:concl} summarizes our main results and discusses some prospects. Throughout this paper we use ``geometrized units'' where $G=c=1$.

\section{Lagrangian and Hamiltonian of a spinning point particle}
\label{sec:hamilton}

In this Section, we review some necessary material for constructing a
Lagrangian and then a Hamiltonian for a spinning point particle in curved spacetime. The
formalism we shall use derives from the early works \cite{HaRe.74,BaIs.75}. 
It has recently been developed in the context of the EFT framework
 \cite{Po.06}. Alternatives and variants to this
formalism can be found in Refs.~\cite{St.al2.08,St.11,Ba.al.09}. The
formalism yields for the equations of motion of spinning particles and
the precession of the spins the classic results valid in general
relativity \cite{Tu.57,Tu.59,Ma.37,Pa.51,CoPa.51,Pi.56,Tr.02,Di.79}.

Let us consider a single spinning point particle moving in a given
curved ``background'' metric $g_{\mu\nu}(x)$. The particle
follows the worldline $r^\mu(s)$, with tangent four-velocity $u^\mu
= \ud r^\mu/\ud s$, where $s$ is a parameter 
along the representative worldline. In a first
stage we do not require that the four-velocity be normalized,
\textit{i.e.}, $s$ need not be the proper time elapsed along the
worldline. To describe some internal degrees of freedom to be
associated with the particle's spin, we introduce a moving tetrad
$e_\alpha^{\phantom{\alpha}\mu}(s)$ along the trajectory. The tetrad is 
orthonormal, in the sense that
$g_{\mu\nu}\,e_\alpha^{\phantom{\alpha}\mu}e_\beta^{\phantom{\beta}\nu}
= \eta_{\alpha\beta}$, and defines a ``body-fixed''
  frame.\footnote{Here $\eta_{\alpha\beta} = \text{diag}(-1,1,1,1)$
  denotes the Minkowski metric. The indices $\mu$, $\nu$, $\cdots$ are
  the usual spacetime covariant indices, while $\alpha$, $\beta$,
  $\cdots$ are the internal Lorentz indices. The inverse (or dual)
  tetrad $e^\alpha_{\phantom{\alpha}\mu}$, defined by
  $e_\alpha^{\phantom{\alpha}\mu}e^\alpha_{\phantom{\alpha}\nu}=\delta^\mu_\nu$,
  satisfies $\eta_{\alpha\beta}
  \,e^\alpha_{\phantom{\alpha}\mu}e^\beta_{\phantom{\beta}\nu} =
  g_{\mu\nu}$. We also have the completeness relation
  $e_\alpha^{\phantom{\alpha}\mu}e^\beta_{\phantom{\beta}\mu}=\delta^\beta_\alpha$.}
The rotation tensor $\Omega^{\mu\nu}$ associated with that tetrad is defined by
\beq\label{rot}
\frac{\uD e_\alpha^{\phantom{\alpha}\mu}}{\ud s} = - \Omega^{\mu\nu} e_{\alpha\nu}\,,
\eeq
where $\uD/\ud s \equiv u^\nu \nabla_\nu$ is the covariant derivative with respect to the parameter $s$ along the worldline. Equivalently, we have
\beq\label{rot2}
\Omega^{\mu\nu}=e^{\alpha\mu}\frac{\uD e_\alpha^{\phantom{\alpha}\nu}}{\ud s}\,.
\eeq
Because of the normalization of the tetrad the rotation tensor is antisymmetric: $\Omega^{\mu\nu}=-\Omega^{\nu\mu}$. 

\subsection{Lagrangian formulation}

Following \cite{HaRe.74,Po.06}, we look for an action for
the spinning particle that is at once: (i) a covariant scalar, (ii)
an internal Lorentz scalar, and (iii) reparametrization-invariant (\textit{i.e.}, its form must be independent of the parameter used to follow the particle's worldline). We shall assume that the dynamical degrees of freedom are the particle's position
$r^\mu$ and the tetrad $e_\alpha^{\phantom{\alpha}\mu}$. We restrict
ourselves to a Lagrangian depending only on the four-velocity $u^\mu$, the rotation
tensor $\Omega^{\mu\nu}$, and the metric $g_{\mu\nu}$. In particular,
this confines the formalism we are using to a pole-dipole model
and to terms linear in the spins.\footnote{Such a model is ``universal'' in the sense that it can be used for black holes as well as neutrons stars. Indeed, the internal structure of the spinning body appears only at the next $\mathcal{O}(S^2)$, \textit{e.g.} through the rotationally induced quadrupole moment.} Thus, the postulated ``particle'' action is of the type
\beq\label{action}
	S_\text{part}\left[r^\mu, e_\alpha^{\phantom{\alpha}\mu}\right] = \int \ud s \, \hat{L}_\text{part}\left(u^\mu,\Omega^{\mu\nu},g_{\mu\nu}\right) . 
\eeq
We cover the Lagrangian $\hat{L}_\text{part}$ with a hat in order to
distinguish it from the usual definition of the Lagrangian,
$L_\text{part} = \hat{L}_\text{part} \, \ud s/\ud t$ (with $t$ the
coordinate time), that we shall use later.

As it is written, depending only on Lorentz scalars, $\hat{L}_\text{part}$ is automatically a Lorentz scalar. By performing an infinitesimal coordinate transformation, one easily sees that the requirement that the Lagrangian be a covariant scalar specifies its dependence on the metric as being (see \textit{e.g.} Ref.~\cite{BaIs.75}) 
\beq\label{scalar}
2 \frac{\partial \hat{L}_\text{part}}{\partial g_{\mu\nu}} = p^\mu u^\nu + S^{\mu}_{\phantom{\mu}\rho}\Omega^{\nu\rho}\,,\eeq
where we have \textit{defined} the conjugate linear momentum $p^\mu$ and the antisymmetric spin tensor $S^{\mu \nu}$ as
\bseq
\label{conjugate}
\begin{align}
p_\mu &\equiv \frac{\partial \hat{L}_\text{part}}{\partial u^\mu}{\bigg|}_{\Omega, g} \,,\\
S_{\mu\nu} &\equiv 2 \frac{\partial \hat{L}_\text{part}}{\partial \Omega^{\mu\nu}}{\bigg|}_{u, g} \,.
\end{align}
\eseq
Note that the right-hand side (RHS) of Eq.~\eqref{scalar} is necessarily symmetric by exchange of the indices $\mu$ and $\nu$. Finally, imposing the invariance of the action \eqref{action} by reparametrization of the worldline, we find that the Lagrangian must be a homogeneous function of degree one in the velocities $u^\mu$ and $\Omega^{\mu\nu}$. Applying Euler's theorem to the function $\hat{L}_\text{part}(u^\mu,\Omega^{\mu\nu})$ immediately gives
\beq\label{Euler}
\hat{L}_\text{part} = p_\mu u^\mu + \frac{1}{2} S_{\mu\nu} \Omega^{\mu\nu} \, ,
\eeq
where the functions $p_\mu(u^\rho,\Omega^{\rho\sigma})$ and
$S_{\mu\nu}(u^\rho,\Omega^{\rho\sigma})$ must be reparametrization
invariant. Note that, at this stage, their explicit expressions are not known. 
They will be specified only once a \textit{spin supplementary condition} (SSC) is 
imposed, as discussed in Sec.~\ref{subsec:H_z} below.

We now investigate the unconstrained variations of the action
\eqref{action} with respect to $e_\alpha^{\phantom{\alpha}\mu},
r^\mu$, and $g^{\mu \nu}$. First, we vary it with respect to the
tetrad $e_\alpha^{\phantom{\alpha}\mu}$ while keeping the position
$r^\mu$ fixed. We must have a way to distinguish intrinsic variations
of the tetrad from those which are induced by a change of the metric $g_{\mu\nu}$. This is done by decomposing the variation
$\delta e_\alpha^{\phantom{\alpha}\nu}$ according to
\beq\label{variation}
\delta e_\alpha^{\phantom{\alpha}\nu} = e_{\alpha\mu} \left( \delta\theta^{\mu\nu} + \frac{1}{2} \delta g^{\mu\nu} \right) ,
\eeq
in which we have introduced the antisymmetric tensor $\delta\theta^{\mu\nu}\equiv e^{\alpha[\mu}\delta e_\alpha^{\phantom{\alpha}\nu]}$, and where the corresponding symmetric part is simply given by the variation of the metric, \textit{i.e.} $e^{\alpha(\mu}\delta e_\alpha^{\phantom{\alpha}\nu)}=\frac{1}{2}\delta g^{\mu\nu}$. Then we can consider the independent variations $\delta\theta^{\mu\nu}$ and $\delta g^{\mu\nu}$. Varying with respect to $\delta\theta^{\mu\nu}$, but holding the metric fixed, gives the equation of spin precession which is found to be 
\beq\label{prec0}
\frac{\uD S_{\mu\nu}}{\ud s} = \Omega_{\mu}^{\phantom{\mu}\rho}S_{\nu\rho} - \Omega_{\nu}^{\phantom{\nu}\rho}S_{\mu\rho}\,,
\eeq
or, alternatively, using the fact that the RHS of Eq.~\eqref{scalar} is symmetric,
\beq\label{prec}
\frac{\uD S_{\mu\nu}}{\ud s} = p_\mu u_\nu - p_\nu u_\mu\,.
\eeq

We next vary with respect to the particle's position $r^\mu$ while holding the tetrad $e_\alpha^{\phantom{\alpha}\mu}$ fixed. Operationally, this means that we have to parallel-transport the tetrad along the displacement vector, \textit{i.e.}, we have to impose 
\beq\label{parallel}
\delta r^\nu \nabla_\nu e_\alpha^{\phantom{\alpha}\mu}=0\,.
\eeq
A simple way to derive the result is to use locally inertial coordinates, such that $\Gamma_{\nu\rho}^\mu = 0$ along the particle's worldline $r^\mu(s)$. Then, Eq.~\eqref{parallel} yields $\delta e_\alpha^{\phantom{\alpha}\mu}=\delta r^\nu \partial_\nu e_\alpha^{\phantom{\alpha}\mu}=-\delta r^\nu \Gamma_{\nu\rho}^\mu e_\alpha^{\phantom{\alpha}\rho}=0$, and the variation gives the well-known Mathisson-Papapetrou \cite{Ma.37,Pa.51,CoPa.51} equation of motion 
\beq\label{EOM}
\frac{\uD p_\mu}{\ud s} = -\frac{1}{2} u^\nu R_{\mu\nu\rho\sigma} S^{\rho\sigma}\,.
\eeq
With more work, Eq.~\eqref{EOM} can also be derived using an arbitrary coordinate system.

Finally, varying with respect to the metric (while keeping $\delta\theta^{\mu\nu}=0$) gives the stress-energy tensor of the spinning particle. We must take into account the scalarity of the action, as imposed by Eq.~\eqref{scalar}, and we obtain the standard result \cite{Tu.57,Tu.59,Ma.37,Pa.51,CoPa.51,Pi.56,Tr.02,Di.79}
\begin{align}\label{Tmunu}
T^{\mu\nu}_\text{part} &= \int \ud s \,p^{(\mu}\,u^{\nu)}\,\frac{\delta^{(4)}
(x-r)}{\sqrt{-g}} \nonumber\\
&- \nabla_\rho \int \ud s\, S^{\rho(\mu}\,u^{\nu)}\,\frac{\delta^{(4)}
(x-r)}{\sqrt{-g}} \, ,
\end{align}
where $\delta^{(4)}(x-r)$ denotes the four-dimensional Dirac function, such that $\int \ud^4 x \, \delta^{(4)}(x) = 1$. It can easily be checked that $\nabla_\nu T_\text{part}^{\mu\nu}=0$ as a consequence of the equation of motion \eqref{EOM} and the equation of spin precession \eqref{prec}.

\subsection{Hamiltonian formulation}

We now want to define a Hamiltonian associated with the Lagrangian \eqref{Euler}. Because of the reparametrization-invariance condition, performing a Legendre transformation with respect to the variables $u^\mu$ and $\Omega^{\mu\nu}$ yields a vanishing result. Different routes are possible. One may add in the action a mass-shell constraint with a Lagrange multiplier and vary the action, keeping the momentum $p_\mu$ as an independent variable. Through various changes of variables and gauge fixings, the action is transformed into an action which possesses Euler-Lagrange equations of the form of Hamilton's equations, such that the canonical (ADM) Hamiltonian can be read off from these equations (see Ref.~\cite{St.11} for a review).

A different strategy, that we shall follow here, is closer to the usual procedure of classical mechanics. It consists of using as internal dynamical variables the six rotational degrees of freedom of Lorentz matrices, and defining the Hamiltonian by means of a $3+1$ split, with a preferred time coordinate, with the usual Legendre transformation \cite{Ba.al.09}. The tetrad $e_\alpha^{\phantom{\alpha}\mu}$ is decomposed into the product of an internal Lorentz transformation $\Lambda^\beta_{\phantom{\beta}\alpha}$ and a reference orthonormal tetrad field $\epsilon_\alpha^{\phantom{\alpha}\mu}(x)$, which is associated with the background metric $g_{\mu\nu}(x)$ and is evaluated at the particle's position, $x^\mu=r^\mu(s)$.\footnote{Such decomposition was already implicit in the variation of the tetrad we performed to derive the spin precession equation \eqref{prec}.} The point is that using as dynamical variables the six internal angles $\phi^a$ of Lorentz matrices $\Lambda^\beta_{\phantom{\beta}\alpha}$ allows us to have usual-looking Euler-Lagrange equations. We thus write\footnote{The indices $a$, $b$ = 1, $\cdots$, 6 label the internal angles $\phi^a$ of Lorentz matrices satisfying $\eta_{\gamma\delta}\,\Lambda^\gamma_{\phantom{\gamma}\alpha}\Lambda^\delta_{\phantom{\delta}\beta}=\eta_{\alpha\beta}$. Since $\epsilon_\alpha^{\phantom{\alpha}\mu}$ is a tetrad we have $g_{\mu\nu}\,\epsilon_\alpha^{\phantom{\alpha}\mu}\epsilon_\beta^{\phantom{\beta}\nu} = \eta_{\alpha\beta}$ and, of course, still $g_{\mu\nu}\,e_\alpha^{\phantom{\alpha}\mu}e_\beta^{\phantom{\beta}\nu} = \eta_{\alpha\beta}$.}
\beq\label{eeps}
e_\alpha^{\phantom{\alpha}\mu}(\phi,r) = \Lambda^\beta_{\phantom{\beta}\alpha} (\phi)\,\epsilon_\beta^{\phantom{\beta}\mu}(r)\, .
\eeq

Next, we perform a $3+1$ decomposition of all the fields. In particular, we split the particle's position and coordinate velocity according to $r^\mu=(t, \mathbf{r})$ and $v^\mu=(1, \mathbf{v})$, where $\mathbf{v}\equiv\ud \mathbf{r}/\ud t$ and boldface letters denote ordinary spatial vectors (often also denoted with spatial indices $i$, $j$, $\cdots$ = 1, 2, 3). We choose for the parameter $s$ along the worldline the proper time $\tau$ and define the so-called redshift variable $z$ by \cite{De.08,Le.al.12}
\beq\label{z}
z \equiv \frac{\ud\tau}{\ud t} = {\bigl[ - g_{\mu\nu}(t, \mathbf{r}) \,v^\mu v^\nu \bigr]}^{1/2} .
\eeq
Then the ordinary Lagrangian $L_\text{part} \equiv z \hat{L}_\text{part}$, such that $S_\text{part} = \int \ud t \, L_\text{part}$, becomes a function of the variables $t$, $\mathbf{r}$, $\mathbf{v}$, $\phi^a$, and the ordinary derivative $\dot{\phi}^a\equiv\ud \phi^a/\ud t$, with the dependence in $t$ and $\mathbf{r}$ entering only through the background metric or, rather, through the reference tetrad evaluated at the particle's location:
\beq\label{Lpart}
L_\text{part}=L_\text{part}\bigl[\mathbf{v},\phi^a,\dot{\phi}^a;\epsilon_\alpha^{\phantom{\alpha}\mu}(t,\mathbf{r})\bigr]\, .
\eeq
We have ordinary Euler-Lagrange equations for the generalized coordinates $( \mathbf{r}, \phi^a )$ of the spinning particle (spatial vectors ranging over three components $i = 1, 2, 3$ while the rotation label runs over $a = 1, \cdots, 6$):
\bseq\label{Lagrange}
\begin{align}
\frac{\ud P_i}{\ud t} &= \frac{\partial L_\text{part}}{\partial r^i} \, , \label{Lag_p} \\
\frac{\ud P_{\phi_a}}{\ud t} &= \frac{\partial L_\text{part}}{\partial \phi^a} \, , \label{Lag_pi}
\end{align}
\eseq
where the conjugate momenta are defined in the usual way by $P_i \equiv \partial L_\text{part}/\partial v^i$ and $P_{\phi_a} \equiv \partial L_\text{part}/\partial \dot{\phi}^a$. They explicitly read
\begin{subequations}\label{momenta}
	\begin{align}
		P_i &= p_i + \frac{1}{2} S_{\mu \nu} \, \epsilon^{\alpha\mu} \nabla_i \epsilon_\alpha^{\phantom{\alpha}\nu} \, , \label{Pi} \\
		P_{\phi_a} &= \frac{1}{2} S_{\alpha\beta} \, \Lambda^{\gamma\beta} \, \frac{\partial \Lambda_\gamma^{\phantom{\gamma}\alpha}}{\partial\phi^a} \, , \label{pia}
	\end{align}
\end{subequations}
where $S_{\alpha\beta} = e_\alpha^{\phantom{\alpha}\mu} e_\beta^{\phantom{\beta}\nu}S_{\mu\nu}$ denote the tetrad components of the spin tensor in the ``body-fixed'' frame. The Euler-Lagrange equations \eqref{Lag_p} and \eqref{Lag_pi} are equivalent to the equations of motion \eqref{EOM} and spin precession \eqref{prec}, respectively \cite{Ba.al.09}. In particular, the spin precession equation is recovered in this formalism as the natural fact that the spin components remain constant in the frame attached to the body:
\beq\label{prec2}
	\frac{\ud S_{\alpha\beta}}{\ud \tau} = 0 \, .
\eeq

Finally, the Hamiltonian is simply given in this approach by the Legendre transformation as
\beq\label{legendre}
H_\text{part} = P_i \, v^i + P_{\phi_a} \dot{\phi}^a - L_\text{part}\, ,
\eeq
and Hamilton's equations yield Eqs.~\eqref{EOM} and \eqref{prec}, or equivalently \eqref{prec2} \cite{Ba.al.09}. The Hamiltonian is a function of conjugate variables and again depends on the position $r^\mu$ only through the background tetrad:
\beq\label{Hpart}
H_\text{part}=H_\text{part}\bigl[P_i,\phi^a,P_{\phi_a};\epsilon_\alpha^{\phantom{\alpha}\mu}(t,\mathbf{r})\bigr]\, .
\eeq
An explicit computation of the Hamiltonian \eqref{legendre} with the help of Eqs.~\eqref{Euler} and \eqref{momenta} yields \cite{Ba.al.09}
\beq\label{hamilton}
H_\text{part} = - p_t - \frac{1}{2} S_{\mu\nu} \,\epsilon^{\alpha\mu} \nabla_t \epsilon_\alpha^{\phantom{\alpha}\nu} \, .
\eeq
Both terms should be understood as being functions of the conjugate variables, like in Eq.~\eqref{Hpart}. As in ordinary classical mechanics, this is obtained by inverting Eqs.~\eqref{momenta} to obtain $v^i$ and $\dot{\phi}^a$ as functions of $r^i$, $P_i$, $\phi^a$ and $P_{\phi_a}$. However, here we need an additional relation to express $p_t$ as a function of the conjugate variables; this is provided by Eq.~\eqref{Euler} which, in the $3+1$ split, gives $p_t = L_\text{part} - v^i p_i - \frac{z}{2} S_{\mu\nu}\Omega^{\mu\nu}$.

\subsection{Relating the Hamiltonian to the redshift}\label{subsec:H_z}

In this Section we shall relate the Lagrangian $L_\text{part}$ and the Hamiltonian $H_\text{part}$ to the redshift variable $z$. To do so we must introduce a realistic physical model for the spin of the point particle in the pole-dipole approximation.

Up to now we have considered unconstrained variations of the action \eqref{action}, which describes the particle's internal degrees of freedom by the 6 independent components of the tetrad $e_\alpha^{\phantom{\alpha}\mu}$ --- a $4\times 4$ matrix subject to the 10 constraints $g_{\mu\nu}\,e_\alpha^{\phantom{\alpha}\mu}e_\beta^{\phantom{\beta}\nu} = \eta_{\alpha\beta}$ --- or equivalently by the 6 internal angles $\phi^a$. To correctly account for the number of degrees of freedom associated with the spin, we must impose 3 SSC. In this Section, we adopt the Tulczyjew covariant conditions \cite{Tu.57,Tu.59}
\beq\label{SSC}
S^{\mu\nu} p_\nu = 0 \, .
\eeq
It would be possible to specify the Lagrangian in \eqref{action} so that the constraints \eqref{SSC} are directly the consequence of the equations derived from that Lagrangian \cite{HaRe.74}. Alternatively, one could also introduce Lagrange multipliers into the action to enforce these constraints \cite{St.11}. Here, for the sake of simplicity, we shall rather impose the constraints \eqref{SSC} directly in the space of solutions. Furthermore, by contracting Eq.~\eqref{prec} with $p^\nu$ and using the equation of motion \eqref{EOM}, one obtains the relation linking the four-momentum $p_\mu$ to the four-velocity $u_\mu$ as\footnote{We denote $(pu)\equiv p_\nu u^\nu$. By further contracting Eq.~\eqref{pu} with $u^\mu$ we obtain an explicit expression for $(pu)$, which can then be substituted back into \eqref{pu}.}
\beq\label{pu}
p_\mu (pu) + m^2 u_\mu = \frac{1}{2} u^\lambda R^\nu_{\phantom{\nu} \lambda \rho \sigma} S^{\rho\sigma}S_{\mu \nu}\, ,
\eeq
where we have defined $m^2 \equiv - g^{\mu\nu} p_\mu p_\nu$. The parameter $m$ is the mass of the particle, and it can be checked using Eqs.~\eqref{SSC} and \eqref{pu} that it is constant along the trajectory, that is $\ud m/\ud \tau=0$.

Henceforth we restrict our attention to spin-orbit (SO) interactions, which are \textit{linear} in the spins. Neglecting quadratic spin-spin (SS) and higher-order interactions, the linear momentum is simply proportional to the four-velocity: $p_\mu=m u_\mu + \mathcal{O}(S^2)$. To linear order, Eq.~\eqref{prec} reduces to the equation of parallel transport for the spin tensor, $\uD S_{\mu\nu} / \ud \tau = \mathcal{O}(S^2)$, and the corresponding stress-energy tensor reads
\begin{align}\label{Tmunupart}
	T^{\mu\nu}_\text{part} &= m\,\frac{v^\mu v^\nu}{z} \, \frac{\delta^{(3)}(\mathbf{x}-\mathbf{r})}{\sqrt{-g}} \nonumber\\ &- \nabla_\rho \biggl( S^{\rho(\mu} v^{\nu)} \, \frac{\delta^{(3)}(\mathbf{x}-\mathbf{r})}{\sqrt{-g}}\biggr) + \mathcal{O}(S^2) \,,
\end{align}
where $\delta^{(3)}(\mathbf{x}-\mathbf{r})$ is the three-dimensional Dirac function, such that $\int \ud^3 x \, \delta^{(3)}(\mathbf{x}) = 1$, and we use the parametrization by the proper time $\tau$ and the notation \eqref{z}, with $v^\mu=z\,u^\mu$. Furthermore, thanks to \eqref{Euler} the Lagrangian at linear order in the spin reads
\beq\label{LSO}
L_\text{part} = z \left( - m + \frac{1}{2} S_{\mu\nu}\Omega^{\mu\nu} \right) + \mathcal{O}(S^2)\,.
\eeq

We now want to compute from Eq.~\eqref{LSO} the partial derivative of the Lagrangian with respect to the mass $m$ of the particle, while holding the dynamical degrees of freedom $\mathbf{r}$, $\mathbf{v}$, $\phi^a$, and $\dot{\phi}^a$ fixed. The first term in \eqref{LSO} will contribute in an obvious way; however we must also control the partial derivative of the spin tensor $S_{\mu\nu}$, which is the conjugate of the rotation tensor $\Omega^{\mu\nu}$, with respect to the mass. Let us show that, in fact, the second term in Eq.~\eqref{LSO} will not contribute to the result at the required $\mathcal{O}(S)$.

To show this we choose the time-like tetrad vector to agree with the four-momentum rescaled by the mass, \textit{i.e.} $e_0^{\phantom{0}\mu} = p^\mu / m$. Then, using Eq.~\eqref{rot} and the equation of motion \eqref{EOM} we get $\Omega^{\mu\nu}p_\nu = \frac{1}{2} u^\nu R^\mu_{\phantom{\mu}\nu\rho\sigma} S^{\rho\sigma}$. Comparing with the SSC \eqref{SSC} and using $p_\mu=m u_\mu + \mathcal{O}(S^2)$, we infer that $S_{\mu\nu}$ and $\Omega_{\mu\nu}$ must be proportional, up to a small curvature coupling to the spin and some higher-order spin terms: $S_{\mu\nu} \propto \Omega_{\mu\nu} - \frac{1}{2m}R_{\mu\nu\rho\sigma} S^{\rho\sigma} + \mathcal{O}(S^3)$. For instance, such a proportionality relation is verified in models for the relativistic spherical top \cite{HaRe.74,Po.06}, and the constant of proportionality is associated with the moment of inertia, say $I$, of the spherical top. The previous relation can be solved iteratively, yielding an expression for $S_{\mu\nu}$ as a function of $m$, $I$, $R_{\mu\nu\rho\sigma}$, and $\Omega^{\mu\nu}$, which schematically reads $S \sim I \big[ \Omega - \frac{I}{2m} R \Omega + {\left(\frac{I}{2m}\right)}^2 R R \Omega + \cdots \big]$. The functional dependence on $m$ is explicit in this expression. Taking a partial derivative with respect to the mass, and replacing back $\Omega^{\mu\nu}$ in terms of the spin, we deduce that $\partial S_{\mu\nu}/\partial m = \mathcal{O}(S)$. Since the rotation tensor is also $\mathcal{O}(S)$, we finally get $\Omega^{\mu\nu}\partial S_{\mu\nu}/\partial m=\mathcal{O}(S^2)$. We thus conclude that the second term in Eq.~\eqref{LSO} will not contribute at linear order in the spin. Thus, we have proven within the covariant SSC (\ref{SSC}) that
\beq\label{dLdm}
\frac{\partial L_\text{part}}{\partial m} = - z + \mathcal{O}(S^2)\,.
\eeq
Using the properties of the Legendre transformation \eqref{legendre}, we readily find that the partial derivative of the Hamiltonian with respect to the mass, holding the conjugate variables $\mathbf{r}$, $P_i$, $\phi^a$, and $P_{\phi_a}$ fixed, is
\beq\label{dHpartdm}
\frac{\partial H_\text{part}}{\partial m} = \,z + \mathcal{O}(S^2)\,.
\eeq
The latter result will be central to the derivation of the first law of mechanics for binary systems of spinning particles in Sec.~\ref{sec:proof}. For the moment we note that it has been proven only for a single spinning particle moving in a given curved ``background'' spacetime. 

\section{Lagrangian and Hamiltonian of interacting spinning particles}
\label{sec:Fokker}

In this Section, we consider a self-gravitating matter system consisting of $N$ spinning point particles, which we shall label by $A, B = 1, \cdots, N$. The Lagrangian for each particle is of the form \eqref{Lpart} in a given metric, so the system of $N$ particles (interacting only through gravitation) is described by the matter Lagrangian
\beq\label{Lmat}
L_\text{mat} = \sum_{A=1}^N L_\text{part}\bigl[\mathbf{v}_A,\phi^a_A,\dot{\phi}^a_A,m_A;\epsilon_\alpha^{\phantom{\alpha}\mu}(r_A)\bigr]\, .
\eeq
Here we have added for later convenience the dependence on the mass $m_A = (- p_\mu^A\,p_A^{\mu})^{1/2}$. Neglecting terms quadratic or higher order in the spins, this Lagrangian reduces to
\beq\label{LmatSO}
L_\text{mat} = \sum_{A=1}^N z_A \left[- m_A + \frac{1}{2} S^A_{\mu\nu}\Omega_A^{\mu\nu} + \mathcal{O}(S_A^2)\right] .
\eeq
In the above equation, the spins $S^A_{\mu\nu}$ should be seen as functions of $r_A$, $\mathbf{v}_A$, $\phi_A^a$, and $\dot{\phi}_A^a$ through their relations to the rotation tensors $\Omega_A^{\mu\nu}$, as discussed in the previous Section. Next, we add the usual Einstein-Hilbert term for gravitation, $L_\text{EH} = \frac{1}{16\pi} \int \ud^3 x \sqrt{-g} R$, and obtain the total Lagrangian describing the ``matter $+$ gravitation'' system as
\begin{align}\label{Ltot}
L = L_\text{mat}\bigl[\mathbf{v}_A,\phi^a_A,\dot{\phi}^a_A,m_A;\epsilon_\alpha^{\phantom{\alpha}\mu}(r_A)\bigr] + L_\text{EH}\bigl[g_{\mu\nu}\bigr]\,,
\end{align}
in which the spacetime metric reads $g_{\mu\nu}=\epsilon_{\alpha\mu}\,\epsilon^\alpha_{\phantom{\alpha}\nu}$, and the brackets refer, with obvious notations, to a functional dependence. Varying the Lagrangian \eqref{Ltot} with respect to the tetrad $\epsilon_\alpha^{\phantom{\alpha}\mu}$ we get the Einstein field equations $G^{\mu\nu} = 8 \pi \,T^{\mu\nu}_\text{mat}$, where the matter tensor is a sum of terms of the type \eqref{Tmunupart}, one for each particle.

\subsection{Fokker Lagrangian}

We shall define the Fokker Lagrangian \cite{Fo.29} by formally solving the Einstein field equations and eliminating the gravitational field degrees of freedom in Eq.~\eqref{Ltot}. Writing the metric as $g_{\mu\nu}=\eta_{\mu\nu}+h_{\mu\nu}$, the field equations are solved perturbatively in powers of $h_{\mu\nu}$ by using the graviton propagator. As usual we discard in the Einstein-Hilbert action a total derivative and consider instead the Landau-Lifshitz Lagrangian, which involves only first-order derivatives of the metric; symbolically $L_\text{LL} = \frac{1}{16\pi} \int \ud^3 x \sqrt{-g}\,\Gamma\,\Gamma$. Furthermore, we need to specify a coordinate system to solve the field equations; this is done by adding some \textit{gauge-fixing} terms to the action. Choosing the harmonic coordinate system, this yields the ``harmonic-gauge-relaxed'' Lagrangian $L_\text{LL}^\text{harm}$ which contains at quadratic order the graviton propagator $\mathcal{P}_{\mu\nu\rho\sigma}$, say $L_\text{LL}^\text{harm} = -\frac{1}{2} \int \ud^3 x \, [ h\mathcal{P}^{-1}h +\mathcal{O}(h^3)]$ (see \textit{e.g.} App.~A of Ref.~\cite{DaEs.96}). We thus consider
\begin{align}\label{Ltotharm}
L^\text{harm} = L_\text{mat}\bigl[\mathbf{v}_A,\phi^a_A,\dot{\phi}^a_A,m_A;\epsilon_\alpha^{\phantom{\alpha}\mu}(r_A)\bigr] + L_\text{LL}^\text{harm}\bigl[g_{\mu\nu}\bigr]\,.
\end{align}

Next, we perturbatively solve the harmonic-gauge Einstein field equations and we formally re-expand the solution in a PN approximation. Expanding the retardations of retarded integrals in a PN scheme yields accelerations and time derivatives of accelerations. It will also generate higher time derivatives of the spin variables. The solution, valid at any point $x^\mu$ in the near-zone, is a function of all the source parameters,
\beq\label{epsbar}
\bar{\epsilon}_\alpha^{\phantom{\alpha}\mu}(x) \equiv \bar{\epsilon}_\alpha^{\phantom{\alpha}\mu}\bigl[x;\mathbf{r}_B,\mathbf{v}_B, \mathbf{a}_B,\phi^a_B,\dot{\phi}^a_B,\ddot{\phi}^a_B,m_B\bigr]\,,
\eeq
where we collectively denote the higher time derivatives by $\mathbf{a}_B=(\ud\mathbf{v}_B/\ud t, \ud^2\mathbf{v}_B/\ud t^2, \cdots)$ and symbolically by $\ddot{\phi}^a_B$ for the spin variables. Then the near-zone metric solution is $\bar{g}_{\mu\nu}=\bar{\epsilon}_{\alpha\mu}\,\bar{\epsilon}^\alpha_{\phantom{\alpha}\nu}$. Thus, by construction the solution \eqref{epsbar} satisfies
\begin{align}\label{solEE}
\frac{\delta L^\text{harm}}{\delta \bar{\epsilon}_\alpha^{\phantom{\alpha}\mu}} &\equiv \frac{\delta L_\text{mat}}{\delta \epsilon_\alpha^{\phantom{\alpha}\mu}}\bigl[\mathbf{v}_A, \phi^a_A,\dot{\phi}^a_A,m_A; \bar{\epsilon}_\alpha^{\phantom{\alpha}\mu}(r_A)\bigr] \nonumber\\ &+ \frac{\delta L_\text{LL}^\text{harm}}{\delta \epsilon_\alpha^{\phantom{\alpha}\mu}}\bigl[\bar{g}_{\mu\nu}\bigr] = 0 \,.
\end{align}

The Fokker Lagrangian \cite{Fo.29} is now obtained by inserting the formal solution \eqref{epsbar} into \eqref{Ltotharm}; hence 
\begin{align}\label{LFharm}
L_\text{F}^\text{harm} = L_\text{mat}\bigl[\mathbf{v}_A,\phi^a_A,\dot{\phi}^a_A,m_A;\bar{\epsilon}_\alpha^{\phantom{\alpha}\mu}(r_A)\bigr] + L_\text{LL}^\text{harm}\bigl[\bar{g}_{\mu\nu}\bigr]\,.
\end{align}
This Lagrangian gives the correct equations of motion (and spin precession) for the matter variables. Its variational derivative with respect to either the position $\mathbf{r}_A$ or spin parameters $\phi^a_A$ [which we collectively denote as $\bm{\sigma}_A\equiv (\mathbf{r}_A, \phi^a_A)$] reads 
\begin{align}\label{varLF}
\frac{\delta L_\text{F}^\text{harm}}{\delta \bm{\sigma}_A} &= \frac{\delta L_\text{mat}}{\delta \bm{\sigma}_A}\bigl[\mathbf{v}_{B}, \phi^a_{B},\dot{\phi}^a_{B},m_{B}; \bar{\epsilon}_\alpha^{\phantom{\alpha}\mu}(r_{B})\bigr] \nonumber\\ &+ \frac{\delta \bar{\epsilon}_\alpha^{\phantom{\alpha}\mu}}{\delta \bm{\sigma}_A}\,\frac{\delta L^\text{harm}}{\delta \bar{\epsilon}_\alpha^{\phantom{\alpha}\mu}} \,,
\end{align}
where the last term vanishes thanks to Eq.~\eqref{solEE}.\footnote{Notice that in the last term of Eq.~\eqref{varLF} the functional derivative $\delta\bar{\epsilon}_\alpha^{\phantom{\alpha}\mu}/\delta \bm{\sigma}_A$ is to be taken in a generalized sense, because of the presence of accelerations $\mathbf{a}_B$ and $\ddot{\phi}^a_B$.} The equations of motion and precession are precisely those we expect for spinning particles within the background $\bar{\epsilon}_\alpha^{\phantom{\alpha}\mu}$ generated by all the particles themselves, \textit{i.e.}
\beq\label{eomF}
\frac{\delta L_\text{mat}}{\delta \bm{\sigma}_A}\bigl[\mathbf{v}_{B}, \phi^a_{B},\dot{\phi}^a_{B},m_{B}; \bar{\epsilon}_\alpha^{\phantom{\alpha}\mu}(r_{B})\bigr] = 0\,.
\eeq
To obtain this result, the Fokker Lagrangian \eqref{LFharm} must involve not only the matter part evaluated in the background generated by the particles, but also, crucially, the Einstein-Hilbert (or Landau-Lifshitz) part for the gravitational field.

We can now use the properties of the Fokker Lagrangian to compute its partial derivative with respect to one of the masses, say $m_A$, holding all dynamical variables (and the other masses) fixed. As before we obtain 
\begin{align}\label{dLdmA}
\frac{\partial L_\text{F}^\text{harm}}{\partial m_A} &= \frac{\partial L_\text{mat}}{\partial m_A}\bigl[\mathbf{v}_{B}, \phi^a_{B},\dot{\phi}^a_{B},m_{B}; \bar{\epsilon}_\alpha^{\phantom{\alpha}\mu}(r_{B})\bigr] \nonumber\\ &+ \frac{\partial \bar{\epsilon}_\alpha^{\phantom{\alpha}\mu}}{\partial m_A}\,\frac{\delta L^\text{harm}}{\delta \bar{\epsilon}_\alpha^{\phantom{\alpha}\mu}} \,,
\end{align}
where again the last term vanishes thanks to Eq.~\eqref{solEE}, such that it remains only
\beq\label{dLdmA2}
\frac{\partial L_\text{F}^\text{harm}}{\partial m_A} = \frac{\partial L_\text{part}}{\partial m_A}\bigl[\mathbf{v}_{A}, \phi^a_{A},\dot{\phi}^a_{A},m_A; \bar{\epsilon}_\alpha^{\phantom{\alpha}\mu}(r_{A})\bigr] \,,
\eeq
where we used Eq.~\eqref{Lmat} to express the result in terms of the single-particle Lagrangian for particle $A$. From \eqref{LmatSO} and the proof given above Eq.~\eqref{dLdm}, we have, at linear order in the spin $S_A$,
\beq\label{dLdmA3}
\frac{\partial L_\text{F}^\text{harm}}{\partial m_A} = - z_A + \mathcal{O}(S_A^2)\,.
\eeq
We emphasize that this result is valid in the background field generated by the system of $N$ particles; the redshift variable therein reads in full glory
\beq\label{redshiftA}
z_A = \left(- \bar{g}_{\mu\nu}\bigl[r_A;\mathbf{r}_B,\mathbf{v}_B,\mathbf{a}_B,\phi^a_B,\dot{\phi}^a_B,\ddot{\phi}^a_B,m_B\bigr] v_A^\mu v_A^\nu\right)^{1/2} .
\eeq
In particular, it depends on the masses $m_B$ of the other particles in the system.

\subsection{Fokker Hamiltonian}\label{subsec:H_Fokker}

That a relationship analogous to Eq.~\eqref{dLdmA3} holds also for the Hamiltonian is not completely straightforward, because of the presence of higher-order derivatives (notably accelerations) in the harmonic-coordinate Fokker Lagrangian \eqref{LFharm}. The higher-order derivatives come from the replacement of the metric by the perturbative solution \eqref{epsbar}, and from explicit time derivatives in the self-interaction terms (such as $\partial h\,\partial h$) of the graviton action. However, as shown in Ref.~\cite{DaSc.91} (see also \cite{DaSc.85,deA.al.01}), one can remove from a PN expanded Lagrangian the higher-order derivatives of $\mathbf{v}_A$ through a suitable redefinition of the position variables, $\mathbf{r}_A \to \mathbf{r}_A^\text{new}$, and the addition of a total time derivative to the Lagrangian.\footnote{As usual, non-linear contributions in accelerations can be removed by the addition of so-called ``multiple-zero'' terms \cite{BaOC.80,BaOC2.80}.} This process can be implemented to \textit{all} orders in PN theory \cite{DaSc.85,DaSc.91,deA.al.01}. The perturbative method of Ref.~\cite{DaSc.91} is very general, and there should be no obstacle in applying it to the spin variables, \textit{i.e.} in removing from a Lagrangian the higher-order derivatives of $\dot{\phi}^a_A$ through a redefinition $\phi^a_A \to {(\phi^a_A)}^\text{new}$ and the addition of a total time derivative. Thus we can eliminate all the higher-order time derivatives in the harmonic-coordinate Fokker Lagrangian \eqref{LFharm} by introducing the new (non-harmonic) Lagrangian
\beq\label{nonharmF}
	L_\text{F} = L_\text{F}^\text{harm} + \frac{\delta L_\text{F}^\text{harm}}{\delta \mathbf{r}_A}\,\delta \mathbf{r}_A + \frac{\delta L_\text{F}^\text{harm}}{\delta \phi^a_A}\,\delta \phi^a_A + \frac{\ud G}{\ud t}\,,
\eeq
where we denote $\delta\mathbf{r}_A\equiv\mathbf{r}_A^\text{new}-\mathbf{r}_A$ and $\delta\phi^a_A\equiv(\phi^a_A)^\text{new}-\phi^a_A$, and $G$ is some function of time that does not affect the dynamics. The Lagrangian \eqref{nonharmF} now depends on $\mathbf{r}_A^\text{new}$ and $\mathbf{v}_A^\text{new}$, but no longer on accelerations $\mathbf{a}_A^\text{new}$; and it depends on the spins only through $(\phi_A^a)^\text{new}$ and $(\dot{\phi}_A^a)^\text{new}$. Also, the non-harmonic metric (or tetrad) solution of the field equations in the new variables takes the same form as in Eq.~\eqref{epsbar}, but without the contributions from higher time derivatives. This is because the redefinition of the positions $\mathbf{r}_A \to \mathbf{r}_A^\text{new}$ can be seen as being induced by a coordinate transformation of the ``bulk'' near-zone metric, and that metric when evaluated at the location of the particles is necessarily free of higher derivatives (see Sec.~3 in Ref.~\cite{DaSc.85}). Thus, we have the same result as in Eq.~\eqref{dLdmA3}. We shall from now on remove the superscript ``new'' on all variables, but we shall keep in mind that they correspond to non-harmonic coordinates.

Finally, we have an ordinary Lagrangian $L_\text{F}$ depending on the dynamical variables $(\mathbf{r}_A,\mathbf{v}_A,\phi_A^a,\dot{\phi}_A^a)$; like in Eq.~\eqref{legendre} we apply a standard Legendre transformation to obtain the corresponding Hamiltonian, say $H_\text{F}$. Defining $P_i^A \equiv \partial L_\text{F}/\partial v_A^i$ and $P_{\phi_a}^A \equiv \partial L_\text{F}/\partial \dot{\phi}^a_A$, we have
\beq\label{FH}
H_\text{F} = \sum_{A=1}^N \left( P_i^A v_A^i + P_{\phi_a}^A \dot{\phi}_A^a \right) - L_\text{F}\, ,
\eeq
where $\mathbf{v}_A$ and $\dot{\phi}_A^a$ are now viewed as functions of the conjugate variables $\mathbf{r}_B$, $P_i^B$, $\phi_B^a$, and $P_{\phi_a}^B$, obtained by inverting the definitions of the momenta $P_i^A$ and $P_{\phi_a}^A$. And like in Eq.~\eqref{dHpartdm}, a consequence of the Legendre transformation \eqref{FH} is that the partial derivative of the Hamiltonian with respect to the mass $m_A$, holding the conjugate variables $\mathbf{r}_B$, $P_i^B$, $\phi^a_B$, and $P_{\phi_a}^B$ (and the masses $m_B \neq m_A$) fixed, is
\beq\label{dHdm}
\frac{\partial H_\text{F}}{\partial m_A} = \,z_A + \mathcal{O}(S_A^2)\,.
\eeq
The result \eqref{dHdm} is the main ingredient of the first law of mechanics for binary systems of spinning point particles, which is the topic of Sec.~\ref{sec:proof}.

\subsection{Constructing the canonical Hamiltonian}

Working within the ADM Hamiltonian formalism, the authors of Refs.~\cite{St.al2.08,StSc.09} showed that, starting from the unconstrained Hamiltonian depending on the spin variables $S^{\mu \nu}$, it is possible to build the constrained canonical Hamiltonian for gravitationally interacting particles by imposing suitable constraints and gauge conditions, and making a change of variables for the positions, momenta, and spins of the particles. On the other hand, starting from the Hamiltonian \eqref{legendre} which depends on the spin variables $\phi^a$ and $P_{\phi_a}$, the construction of the constrained canonical Hamiltonian for a single spinning particle in a given curved spacetime was explicitly done in Ref.~\cite{Ba.al.09}, thereby extending to curved spacetime the classic work by Hanson and Regge \cite{HaRe.74} for a relativistic top in flat spacetime. Being guaranteed that a constrained, canonical ADM Hamiltonian exists for gravitationally interacting spinning point particles \cite{St.al2.08,StSc.09,St.11}, henceforth we assume that the construction of Ref.~\cite{Ba.al.09} can be extended to interacting particles. Let us then briefly review how the constrained Hamiltonian for a single spinning particle was built in Ref.~\cite{Ba.al.09}.

We start from the unconstrained Hamiltonian for a single spinning particle $H_{\rm part}(\mathbf{r},P_i, \phi^a,P_{\phi_a})$, together with the Poisson bracket operation $\{\cdot\,,\cdot\}$ (see Sec.~III B in Ref.~\cite{Ba.al.09} for the
explicit expressions of the Poisson brackets between the relevant
variables). We consider a set of constraints
$\xi_k=0$ with $k = 1, \cdots, 2M$ (with $M < 9$) such that the matrix
$C_{k l} = \{\xi_k,\xi_l\}$ is not singular. Following
Ref.~\cite{HaRe.74}, these constraints can be imposed by (i) replacing
the original Poisson brackets with Dirac brackets,\footnote{For two
  phase-space functions $A$ and $B$, the Dirac brackets are given by
  ${\{A,B\}}_{\rm D} = \{A,B\}+\{A,\xi_k\}\{B,\xi_l\}{[C^{-1}]}_{kl}$, and
  represent the projection of the original symplectic structure onto
  the phase-space surface defined by the constraints.} and (ii)
inserting the constraints directly in the original, unconstrained
Hamiltonian. More specifically, Ref.~\cite{Ba.al.09} first imposed the
Newton-Wigner SSC \cite{HaRe.74}, suitably generalized to curved
spacetime, to eliminate 3 out of the 6 variables $P_{\phi_a}$
($a = 1, \cdots, 6$), and an additional constraint to eliminate 3 out of the 6 variables $\phi^a$, such that the constraint
hypersurface contains the same number of configuration coordinates and
conjugate momenta. Then, the constrained Hamiltonian
$H_C(\mathbf{r},\mathbf{p},\mathbf{S})$ and the phase-space
algebra of the constrained system were computed, showing that they are
canonical at linear order in the particle's spin.
Thus, the variables $\mathbf{r}$, $\mathbf{p}$, and
$\mathbf{S}$---with ${S}^i = \frac{1}{2} \varepsilon^{ijk}
S_{jk}$ where $S_{jk} = \epsilon_j^{\phantom{j}\mu}
\epsilon_k^{\phantom{k}\nu} S_{\mu \nu}$ ($i,j,k=1,2,3$)---obey the
standard commutation relations $\{ r^i, p_j \} = 
\delta^i_j$ and $\{ S^i, S^j \} = \varepsilon^{ijk}
S^k$, all other brackets vanishing. As usual, this algebra can be used to compute the dynamical
evolution of any (time-independent) function $f$ of the canonical
variables according to $\dot{f} \equiv \ud f / \ud t = \{f, H\}$.
Finally, the constrained canonical Hamiltonian constructed in
Ref.~\cite{Ba.al.09} coincides with the ADM Hamiltonian
for a spinning particle moving in a given background.

As said, we shall assume that the construction above can be extended to gravitationally interacting spinning point particles. The resulting constrained Hamiltonian $H_C(\mathbf{r}_A,\mathbf{p}_A,\mathbf{S}_A)$ with $A = 1, \cdots, N$ with usual canonical variables will coincide with the ADM canonical Hamiltonian. Furthermore, being constructed from the class of Fokker Hamiltonians \eqref{FH}, it necessarily satisfies Eq.~\eqref{dHdm}. Thereafter, we restrict ourselves to a binary system of spinning point masses, \textit{i.e.} $A = 1,2$, and we denote the canonical ADM Hamiltonian by $H$.

\section{First law of mechanics for binary point particles with spins}
\label{sec:proof}

\subsection{Derivation of the first law}

In Section \ref{subsec:H_Fokker} we proved the remarkable relationship \eqref{dHdm}, valid when neglecting non-linear terms in the spins of the type $\mathcal{O}(S_1^2)$ and $\mathcal{O}(S_2^2)$, \textit{i.e.}
\beq\label{zA}
\frac{\partial H}{\partial m_A} = z_A \, .
\eeq
Note that this relation is not only valid for linear contributions $\mathcal{O}(S_1)$ and $\mathcal{O}(S_2)$, but also for non-linear spin contributions of the type $\mathcal{O}(S_1 S_2)$. Here the redshift quantity
$z_A$ is defined in terms of the coordinate four-velocity $v_A^\mu =
(1,\mathbf{v}_A)$ of particle $A$, in ADM-type coordinates, by
\beq\label{zAdef}
z_A = \frac{\ud\tau_A}{\ud t} = \Bigl(- \bar{g}_{\mu\nu}\bigl[r_A;\mathbf{r}_B,\mathbf{v}_B,\mathbf{S}_B,m_B\bigr] v_A^\mu v_A^\nu\Bigr)^{1/2}\, ,
\eeq
with $\bar{g}_{\mu\nu}$ being the background metric generated by the two particles, and evaluated at the location of particle $A$. Note that the particle velocity $\mathbf{v}_A$ in the RHS of Eq.~\eqref{zAdef} should be viewed as a function of the conjugate variables $\mathbf{r}_A$, $\mathbf{p}_A$, $\mathbf{S}_A$, obtained by inverting the definition of the momentum $\mathbf{p}_A$. We emphasize that Eqs.~(\ref{zA})--(\ref{zAdef}) are valid for \textit{generic} orbits and spin configurations. For circular orbits and spins aligned or anti-aligned with respect to the orbital angular momentum, $z_A$ would coincide with the (gauge-invariant) redshift observable introduced by Detweiler \cite{De.08}. This is the redshift of a photon emitted from the particle and observed at a large distance, along the symmetry axis $\hat{\mathbf{z}}$ perpendicular to the orbital plane. 

In order to derive the first law, we reduce the ADM Hamiltonian $H(\mathbf{r}_A,\mathbf{p}_A,\mathbf{S}_A,m_A)$, given in a generic frame of reference, to the center-of-mass Hamiltonian, say $H(\mathbf{r},\mathbf{p},\mathbf{S}_A,m_A)$. This is achieved by first imposing that the ADM linear momentum $\mathbf{P}_{\!\text{ADM}} = \mathbf{p}_1 + \mathbf{p}_2$ vanishes, then substituting the individual momenta $\mathbf{p}_A$ for the relative linear momentum $\mathbf{p} \equiv \mathbf{p}_1 = -\mathbf{p}_2$, and denoting the coordinate separation by $\mathbf{r} =  \mathbf{r}_1 - \mathbf{r}_2$. 

We shall also limit our study to spins aligned or anti-aligned with the orbital angular momentum $\mathbf{L}$, and to circular orbits. Defining $\mathbf{L} = L \, \hat{\mathbf{z}}$, where $\hat{\mathbf{z}}$ is the unit vector orthogonal to the orbital plane, we have $\mathbf{S}_A = S_A \, \hat{\mathbf{z}}$, with $\vert S_A \vert < m_A^2$. Using polar coordinates $(r, \varphi)$ in the orbital plane, we can express the Hamiltonian as a function of the separation $r = \vert \mathbf{r} \vert$, of the radial momentum $p_r$, of the azimuthal momentum $p_\varphi = L$, and of the masses $m_A$ and amplitudes $S_A$ of the spins of the two point particles. An unconstrained variation of the Hamiltonian $H(r,p_r,p_\varphi,S_A,m_A)$ thus gives 
\begin{align}\label{dH}
	\delta H &= \frac{\partial H}{\partial r} \, \delta r + \frac{\partial H}{\partial p_r} \, \delta p_r + \frac{\partial H}{\partial L} \, \delta L \nonumber \\
&  + \sum_A \left( \frac{\partial H}{\partial m_A} \, \delta m_A + \frac{\partial H}{\partial S_A} \, \delta S_A \right) ,
\end{align}
where to ease the notation we have omitted to indicate that the partial derivatives with respect to $r$, $p_r$, $L$, ${S}_A$, and $m_A$ are computed keeping all the other variables fixed. If the variation compares one solution of the Hamiltonian dynamics to a neighboring solution, then Hamilton's equations of motion must be satisfied. For circular orbits, this yields $\partial H / \partial r = - \dot{p}_r = 0$, as well as $\partial H / \partial p_r = \dot{r} = 0$. The constant circular-orbit frequency is given by $\partial H / \partial L = \dot{\varphi} = \Omega$, and the Hamiltonian is numerically equal to the ADM mass: $H=M$. Therefore, ``on shell'' we have
\beq\label{dM0}
	\delta M = \Omega \, \delta L + \sum_A \left( \frac{\partial H}{\partial m_A} \, \delta m_A + \frac{\partial H}{\partial S_A} \, \delta S_A \right) .
\eeq

At linear order in the spins, the Hamiltonian can be written as the sum of an orbital part $H_\text{O}$, which does not depend on the spins $S_A$, and a spin part $H_\text{S}=\sum_A \Omega_A S_A$. The partial derivatives of the Hamiltonian with respect to the spins,
\beq\label{OmA}
	\Omega_A = \frac{\partial H}{\partial S_A} \,,
\eeq
are the so-called \textit{precession} frequencies of the spins. Indeed, using the algebra satisfied by the canonical variables, the spins $\mathbf{S}_A$ can easily be seen to satisfy --- in the most general, precessing case --- the Newtonian-looking (but exact) precession equations $\dot{\mathbf{S}}_A = \mathbf{\Omega}_A \times \mathbf{S}_A$. Thus, the usual Euclidean norms of the canonical spins are conserved, $\mathbf{S}_A \cdot \mathbf{S}_A = \text{const}$, and $\mathbf{S}_A$ are also referred to as the constant-in-magnitude spins, or \textit{constant} spins.\footnote{Note that the constant spins are not uniquely defined, because a local Euclidean rotation leaves the magnitude $\mathbf{S}_A \cdot \mathbf{S}_A$ of the spin $\mathbf{S}_A$ unchanged. However, for spins aligned or anti-aligned with the orbital angular momentum, this remaining gauge freedom is irrelevant, as it does not affect the (algebraic) magnitude $S_A$.}

Finally, combining Eqs.~\eqref{zA}, \eqref{dM0}, and \eqref{OmA}, the first law of mechanics takes on the simple form\footnote{A similar result was previously established in \cite{Da.al.02}, see Eq.~(24) there, although without the crucial mass variation terms $z_A \, \delta m_A$ in the right-hand side.}
\beq\label{1st_law_simple}
	\delta M - \Omega \, \delta L = \sum_A \bigl( z_A \, \delta m_A + \Omega_A \, \delta S_A \bigr) \,.
\eeq
This differential relation gives the changes in the ADM mass $M$ and orbital angular momentum $L$ of the \textit{binary} system under small changes in the \textit{individual} masses $m_A$ and spins $S_A$ of the two point particles. We emphasize that Eq.~\eqref{1st_law_simple} is valid for any spin magnitude $\vert S_A \vert < m_A^2$.

It is convenient to replace the orbital angular momentum $L$ in favor of the total (ADM-like) angular momentum $J$. For aligned or anti-aligned spins, we simply have
\beq
\label{J}
J = L + \sum_A S_A \, .
\eeq
In terms of $J$, the first law \eqref{1st_law_simple} reads
\beq
\label{1st_law}
\delta M - \Omega \, \delta J = \sum_A \Bigl[ z_A \, \delta m_A + \left( \Omega_A -  \Omega \right) \delta S_A \Bigr] \,.
\eeq
We recall that the expressions \eqref{1st_law_simple} and \eqref{1st_law} of the first law are, in principle, valid only at linear order in the amplitudes $S_A$ of the constant (or canonical) spin variables $\mathbf{S}_A$. It is true that the definitions \eqref{OmA} of the precession frequencies do not require that the Hamiltonian $H$ be linear in the spins; if contributions quadratic (or higher) in the spins were included in the Hamiltonian, then the precession frequencies $\Omega_A$ would become functions of the spins $S_A$. Nevertheless, our proof in Sec.~\ref{sec:Fokker} of the result \eqref{dHdm} is only valid at linear order in the spins. 

\subsection{Consequences of the first law}
\label{sec:consequences}

We now explore some consequences of the first law for spinning point particles. In particular, we establish algebraic expressions which can be regarded as first integrals associated with the variational relations \eqref{1st_law_simple} and \eqref{1st_law}. 

For two spinning point particles with masses $m_A$ and spins $S_A$, on a circular orbit with azimuthal frequency $\Omega$, the ADM mass $M$, orbital angular momentum $L$, redshifts $z_A$ and precession frequencies $\Omega_A$ are all functions of the five independent variables $(\Omega,m_1,m_2,S_1,S_2)$. Therefore, the first law \eqref{1st_law_simple} is equivalent to the following set of partial differential equations:
\bseq\label{PDE}
	\begin{align}
		\frac{\partial M}{\partial \Omega} - \Omega \frac{\partial L}{\partial \Omega} &= 0 \, , \label{PDE_Omega} \\
		\frac{\partial M}{\partial m_A} - \Omega \frac{\partial L}{\partial m_A} &= z_A \, , \label{PDE_mA} \\
		\frac{\partial M}{\partial S_A} - \Omega \frac{\partial L}{\partial S_A} &= \Omega_A \, . \label{PDE_SA}
	\end{align}
\eseq
Note that the relations \eqref{PDE_Omega} and \eqref{PDE_mA} are unchanged with respect to the results in the non-spinning case [see Eqs.~(2.40) and (2.41) of Paper I], where they have been checked up to 3PN order plus the logarithmic contributions at 4PN and 5PN order. However, in the spinning case we have the additional equations \eqref{PDE_SA} relating the precession frequencies $\Omega_A$ to the partial derivatives of $M$ and $L$ with respect to the spins $S_A$.

By combining Eqs.~\eqref{PDE_Omega}--\eqref{PDE_mA} with numerical calculations of the GSF effect on the redshift $z_1$ of a non-spinning point particle on a circular orbit around a Schwarzschild black hole \cite{De.08,Sa.al.08,Sh.al.11}, the \textit{exact} expressions of the total mass $M$ and orbital angular momentum $L$ could be determined, at leading order beyond the test-particle approximation \cite{Le.al2.12}. This result was used in \cite{Ba.al.12} (see also Ref.~\cite{Ak.al.12}) to improve the knowledge of the EOB model \cite{BuDa.99,BuDa.00,Da.al3.00} for non-spinning binaries. We leave to future work the generalization of these findings to spinning point masses.

By commutation of partial derivatives, Eqs.~\eqref{PDE_mA} and \eqref{PDE_SA} can be combined to give the interesting relation
\beq
\label{zA_OmegaB}
\frac{\partial z_A}{\partial S_B} = \frac{\partial \Omega_B}{\partial m_A} \, .
\eeq
This equation relates the variation in the redshift $z_A$ of particle $A$ under a small change in the spin $S_B$ of particle $B$ to the variation of the precession frequency $\Omega_B$ of particle $B$ under a small change in the mass $m_A$ of particle $A$. Therefore, for $A \neq B$, Eq.~\eqref{zA_OmegaB} reflects some equilibrium state of the spinning point particles under their mutual gravitational attraction. For $A = B$, we obtain a non-trivial relation between the redshift (coinciding, in adapted coordinates, with the helical Killing energy) and the precession frequency of each spinning point particle.

We now derive the first integral associated with the variational first law \eqref{1st_law_simple} by following the same steps as in the proof given in the non-spinning case in Paper I. First, we introduce the convenient combination
\beq
\mathcal{M} \equiv M - \Omega L \, .
\eeq
Making the change of variables $(\Omega,m_1,m_2,S_1,S_2)
\rightarrow (\Omega,m,\nu,\chi_1,\chi_2)$, where $m \equiv m_1 + m_2$
is the total mass, $\nu \equiv m_1 m_2 / m^2$ the symmetric mass
ratio, and $\chi_A \equiv S_A / m_A^2$ the dimensionless spins,
Eqs.~\eqref{PDE_mA}--\eqref{PDE_SA} can be combined to give
\begin{align}
\label{sum}
\sum_A \bigl( m_A z_A + 2 \Omega_A S_A \bigr) &= \sum_A \left( m_A \frac{\partial \mathcal{M}}{\partial m_A} + 2 S_A \frac{\partial \mathcal{M}}{\partial S_A} \right) \nonumber \\
&= m \,\frac{\partial \mathcal{M}}{\partial m} \, . 
\end{align}
Next, we notice that the ratio $\mathcal{M}/m$ is dimensionless and symmetric by the exchange $m_1 \leftrightarrow m_2$ of the particles; it must thus be a function of $m\Omega$, $\nu$, $\chi_1$, and $\chi_2$; see \textit{e.g.} Eqs.~\eqref{Ec_Lc_z1c} below. This last observation implies the relationship $m \, \partial (\mathcal{M}/m) / \partial m = \Omega \, \partial (\mathcal{M}/m) / \partial \Omega$, which when combined with \eqref{PDE_Omega} and \eqref{sum} yields the first integral relation
\beq\label{1st_integral_bis}
M - 2 \Omega L = \sum_A \bigl( m_A z_A + 2 \Omega_A S_A \bigr) \,.
\eeq
Alternatively, Euler's theorem for homogeneous functions provides a more straightforward proof of the result \eqref{1st_integral_bis}. Since the Einstein field equations do not contain any privileged mass scale, the ADM mass must be a homogeneous function of degree one in $L^{1/2}$, $m_A$ and $S_A^{1/2}$. Hence, Euler's theorem implies
\begin{align}
	M(L,m_A,S_A) &= L^{1/2} \frac{\partial M}{\partial L^{1/2}} \nonumber \\ &+ \sum_A \biggl( m_A \frac{\partial M}{\partial m_A} + S_A^{1/2} \frac{\partial M}{\partial S_A^{1/2}} \biggr) \, ,
\end{align}
which when combined with the first law \eqref{1st_law_simple} immediately yields the first integral relation \eqref{1st_integral_bis}. This result generalizes to spinning point particles the first integral relation derived in the non-spinning case [see Eq.~(1.2) of Paper I], as it involves appropriate additional spin terms.

Furthermore, we can use the previous results to obtain a relationship between the spin contributions to the ADM mass $M$ and the sum $\sum_A m_A z_A$ of the redshifted masses. Indeed, Eqs.~\eqref{zA_OmegaB} and \eqref{PDE_SA} successively imply
\beq 
\frac{\partial}{\partial S_B} \sum_A m_A z_A = \sum_A m_A
\frac{\partial \Omega_B}{\partial m_A} = \frac{\partial}{\partial S_B}
\sum_A m_A \frac{\partial \mathcal{M}}{\partial m_A} \, .  
\eeq
By combining Eqs.~\eqref{sum} and \eqref{1st_integral_bis}, we can easily replace the sum $\sum_A m_A \partial \mathcal{M} / \partial m_A$ in the RHS by the expression $- M + 2 \mathcal{M} - 2 \sum_A S_A \partial \mathcal{M} / \partial S_A$. At linear order in the spins, $\partial^2 \mathcal{M} / (\partial S_A \partial S_B) = 0$, and we find
\beq
\label{M_mz_SO}
\frac{\partial}{\partial S_B} \bigl( M + m_1 z_1 + m_1 z_2 \bigr) =  0 \, .
\eeq
This simple relation immediately gives the SO contributions to the ADM mass $M$ once those in the redshift observables $z_1$ and $z_2$ are known, and \textit{vice versa}.

Until now, all results were given in terms of the orbital angular momentum $L$. We can rewrite them using the total angular momentum \eqref{J} instead. The partial differential equations \eqref{PDE} then become
\bseq
	\begin{align}
		\frac{\partial M}{\partial \Omega} - \Omega \frac{\partial J}{\partial \Omega} &= 0 \, , \label{PDE_Omega_J} \\
		\frac{\partial M}{\partial m_A} - \Omega \frac{\partial J}{\partial m_A} &= z_A \, , \label{PDE_mA_J} \\
		\frac{\partial M}{\partial S_A} - \Omega \frac{\partial J}{\partial S_A} &= \Omega_A - \Omega \, . \label{PDE_SA_J}
	\end{align}
\eseq
Equation \eqref{PDE_Omega_J} is the ``thermodynamical'' relation commonly used in PN theory for quasi-circular orbits (see, \textit{e.g.}, Refs.~\cite{Da.al.00,Bl.02}), or in the construction of sequences of quasi-equilibrium initial data for binary black holes and binary neutrons stars \cite{Go.al.02,Gr.al.02,Sh.al.04,Ca.al2.06}. In terms of the total angular momentum, Eq.~\eqref{1st_integral_bis} becomes
\beq\label{1st_integral_ter}
	M - 2 \Omega J = \sum_A \Bigl[ m_A z_A - 2 (\Omega - \Omega_A) S_A \Bigr] \,,
\eeq
which is, as expected, the first integral associated with the variational first law \eqref{1st_law}. The existence of such a simple, linear, algebraic relation between the local quantities $z_A$ and $\Omega_A$ on one hand, and the global quantities $M$ and $J$ on the other hand, is noticeable.

\section{Spin-orbit effects in the particle's redshift observable}
\label{sec:PN}

We employ the ADM Hamiltonian to derive the spin-orbit terms at next-to-leading 2.5PN order in the particle's redshift observable (or helical Killing energy) using the relation \eqref{zA}. We keep neglecting all terms $\mathcal{O}(S_A^2)$ or higher. We do not include non-linear spin interactions of the type $\mathcal{O}(S_1 S_2)$ either, even though they must satisfy our first law of mechanics. The center-of-mass ADM Hamiltonian can then be written as 
\beq
H=H(\mathbf{r},\mathbf{p},\mathbf{S}_A,m_A)=H_\textrm{O}+H_\textrm{SO} + H_{\rm NL SO}\,,
\eeq
where $H_\textrm{O}$ is the orbital (or non-spinning) Hamiltonian, which is known through 3PN order, while the leading-order 1.5PN spin-orbit term reads \cite{DaSc.88}
\beq\label{H_SO}
H_\textrm{SO} = \frac{2}{r^3} \,\mathbf{S}_\textrm{eff} \cdot \mathbf{L}\,,
\eeq
with $\mathbf{L} = \mathbf{r} \times \mathbf{p}$ the orbital angular momentum, $\mathbf{n} = \mathbf{r}/r$ the unit vector pointing from $m_2$ to $m_1$, and 
\beq
\mathbf{S}_\textrm{eff} \equiv \left(1+\frac{3m_2}{4m_1} \right)\mathbf{S}_1+\left(1+\frac{3m_1}{4m_2} \right)\mathbf{S}_2 \, .
\eeq
The next-to-leading 2.5PN SO terms, computed first in the harmonic-coordinates approach \cite{Ta.al.01,Fa.al.06}, and then in the ADM Hamiltonian \cite{Da.al.08}, read\footnote{Note that spin-orbit terms in the ADM Hamiltonian are also known at 3.5PN order \cite{HaSt.11,HaSt2.11}, but we do not include them here.}
\begin{align}\label{H_NLSO}
H_{\rm NL SO} &= \frac{(\mathbf{L} \cdot \mathbf{S}_1)}{r^3}\,\left \{
\left ( - 6 m_1 - 13 m_2 - \frac{5m_2^2}{m_1} \right )\,\frac{1}{r} \right.  \nonumber\\
& \quad \left. + \left [ \left (  \frac{3}{4}\frac{1}{m_1^2} + \frac{3}{2}\frac{1}{m_1\,m_2} \right ) \,(\mathbf{n} \cdot \mathbf{p})^2 \right. \right. \nonumber \\
& \quad \left. \left. + 
\left ( \frac{3}{4}\frac{1}{m_1^2} + \frac{7}{4}\frac{1}{m_1\,m_2} - 
\frac{5}{8}\frac{m_2}{m_1^3} \right ) \,\mathbf{p}^2\, \right ] \right \} \nonumber \\
& \quad\, + 1 \longleftrightarrow 2\,.
\end{align}

As was done in Sec.~\ref{sec:proof}, we now restrict ourselves to circular orbits ($p_r = \mathbf{n} \cdot \mathbf{p} = 0, \Omega = \partial H/\partial L, \partial H/\partial r = 0$) and spins aligned or anti-aligned with the orbital angular momentum, and use Eq.~\eqref{zA} to compute the spin-orbit coupling through 2.5PN order in the redshift observable $z_A$. Remember that the partial derivative with respect to $m_A$ in Eq.~\eqref{zA} is to be taken at fixed $r$, $L$, $S_A$, and $m_B \neq m_A$. A straightforward calculation gives the following leading 1.5PN SO contributions to the redshift associated with particle $1$:\footnote{Hereafter we assume, without any loss of generality, $m_1 \leqslant m_2$. We denote the reduced mass difference by $\Delta \equiv (m_2 -m_1)/m = \sqrt{1-4\nu}$. The redshift observable $z_2$ of particle 2 can immediately be deduced from $z_1$ by setting $\Delta \longrightarrow - \Delta$ and $\chi_1 \longleftrightarrow \chi_2$.}
\begin{align}\label{z1c}
z_1^\mathrm{SO} &= \biggl [ \left( - \frac{1}{3} + \frac{\Delta}{3} + \frac{2}{3} \nu \right) \nu \, \chi_1 \nonumber \\
&\quad + \left( 1 + \Delta - \frac{17}{6} \nu - \frac{5}{6} \Delta \, \nu + \frac{2}{3} \nu^2 \right) \chi_2 \biggr ]x^{5/2} \,,
\end{align}
and similarly for the next-to-leading 2.5PN SO terms
\begin{align}\label{z1NLSO}
z_1^\mathrm{NLSO} &= \biggl [ \left( \frac{1}{2} - \frac{\Delta}{2} + \frac{19}{18} \nu - \frac{19}{18} \Delta \, \nu - \frac{\nu^2}{9} \right) \nu \, \chi_1 \nonumber \\ 
&\quad + \left( \frac{3}{2} + \frac{3}{2} \Delta - \frac{17}{3} \nu - \frac{8}{3} \Delta \, \nu + \frac{179}{36} \nu^2 \right. \nonumber \\
&\quad \left. + \frac{41}{36} \Delta \, \nu^2 - \frac{\nu^3}{9} \right) \chi_2 \biggr ] \, x^{7/2} \,.
\end{align}
We have introduced the usual frequency-related PN parameter $x \equiv (m\Omega)^{2/3}$ and we recall the notation $\chi_A = S_A / m_A^2$. The SO and NLSO terms found in Eqs.~\eqref{z1c} and \eqref{z1NLSO} agree with the results of Ref.~\cite{Fr.al.12}, obtained from a direct calculation based on the near-zone PN metric; see Eq.~\eqref{redshiftA}. This provides an important, non-trivial test of the validity of the relation \eqref{dHdm} derived in Sec.~\ref{sec:Fokker}.

For consistency we verified that Eqs.~\eqref{PDE}, \eqref{zA_OmegaB}, and \eqref{1st_integral_bis} hold in the spin-orbit sector. We also verified that Eq.~\eqref{M_mz_SO} is satisfied by the spin-orbit terms through 2.5PN order. To check those relations we use the spin-orbit contributions to the binding energy $E \equiv M - m$ and orbital angular momentum: 
\bseq
\label{Ec_Lc_z1c}
\begin{align}
E_{\mathrm{SO}} &= - \frac{m \,\nu \, x}{2} \biggl[ \left( \frac{4}{3} - \frac{4}{3} \Delta - \frac{2}{3} \nu \right) \chi_1 \nonumber \\ &\qquad\quad + \left( \frac{4}{3} + \frac{4}{3} \Delta - \frac{2}{3} \nu \right) \chi_2 \biggr] \, x^{3/2}\,, \\ 
E_{\mathrm{NLSO}} &= - \frac{m \nu \, x}{2} \biggl[ \left( 4 - 4 \Delta - \frac{121}{18} \nu + \frac{31}{18} \Delta \, \nu + \frac{\nu^2}{9} \right) \chi_1 \nonumber \\ &\quad + \left( 4 + 4 \Delta - \frac{121}{18} \nu - \frac{31}{18} \Delta \, \nu + \frac{\nu^2}{9} \right) \chi_2 \biggr] x^{5/2}\,, \\
L_{\mathrm{SO}} &= \frac{m \nu \, x}{\Omega} \biggl[ \left( - \frac{5}{3} + \frac{5}{3} \Delta + \frac{5}{6} \nu \right) \chi_1 \nonumber \\ &\qquad\: + \left( - \frac{5}{3} - \frac{5}{3} \Delta + \frac{5}{6} \nu \right) \chi_2 \biggr] x^{3/2} \,, \\ 
L_{\mathrm{NLSO}} &=  \frac{m \nu \, x}{\Omega} \biggl[ \left( - \frac{7}{2} + \frac{7}{2} \Delta + \frac{847}{144} \nu - \frac{217}{144} \Delta \, \nu - \frac{7}{72} \nu^2 \right) \chi_1 \nonumber \\ &+ \left( - \frac{7}{2} - \frac{7}{2} \Delta + \frac{847}{144} \nu + \frac{217}{144} \Delta \, \nu - \frac{7}{72} \nu^2 \right) \chi_2 \biggr] x^{5/2} \, .
\end{align}
\eseq
We need also the orbital (\textit{i.e.} spin-independent) part of the precession frequencies at next-to-leading order. We compute it from the definition \eqref{OmA} together with the SO Hamiltonian \eqref{H_SO}--\eqref{H_NLSO}, and find
\begin{align}\label{Omega1}
	\Omega_1 &= \Omega \, \biggl\{ \left( \frac{3}{4} + \frac{3}{4} \Delta + \frac{\nu}{2} \right) x \nonumber \\ &+ \left( \frac{9}{16} + \frac{9}{16} \Delta + \frac{5}{4} \nu - \frac{5}{8} \Delta \, \nu - \frac{\nu^2}{24} \right) x^2 \biggr\} \, .
\end{align}
Note that when computing the partial derivatives to prove Eqs.~\eqref{PDE}, \eqref{zA_OmegaB}, and \eqref{M_mz_SO}, we need to hold fixed the spin variables $S_A$ and not the reduced spins $\chi_A$.

We conclude that the redshift observables $z_A$ satisfy the first law of mechanics in the non-spinning sector, up to 3PN order included, as well as for the leading-order 4PN and next-to-leading order 5PN logarithmic terms (Sec.~II D of Paper I), and, in the spin-orbit sector, up to the next-to-leading 2.5PN order.

\section{Binary black holes in corotation}
\label{sec:corot}

In this Section, we consider the particular case of corotating point particles on a circular orbit. By modeling each spinning point particle by an isolated rotating black hole, and comparing with the first law of binary black hole mechanics \cite{Fr.al.02}, we define and compute the proper rotation frequencies of the corotating point particles.

\subsection{Spinning point particles as Kerr black holes}
\label{subsec:corot_Kerr}

In Sec.~\ref{sec:proof} we derived the first law of mechanics for spinning point particles [see Eq.~\eqref{1st_law}]. We now derive an alternative version of this result that can heuristically be applied to binary black holes carrying spins. Clearly, in order to derive a first law for \textit{extended} bodies such as rotating black holes, we must supplement the pole-dipole model we have used so far with some ``constitutive relations'' $m_A(\mu_A,S_A,\cdots)$ specifying the energy content of the bodies, \textit{i.e.} the relations of their masses $m_A$ to the spins $S_A$ and to some ``irreducible'' masses $\mu_A$. In principle, these constitutive relations should also involve other parameters such as some external (tidal) multipole moments due to the environment; in the two-body case, they would thus depend on the orbital separation and the mass of the other body.

By analogy with the first law of mechanics for a single black hole \cite{Ba.al.73}, we \textit{define} for each spinning point particle the analogue of an irreducible mass $\mu_A\equiv m_A^\text{irr}$ \textit{via} the variational relation $\delta m_A = c_A \, \delta \mu_A + \omega_A \, \delta S_A$, in which the ``response coefficients'' $c_A$ and $\omega_A$ are associated with the internal structure:
\beq\label{dmA}
c_A \equiv \frac{\partial m_A}{\partial \mu_A}{\bigg|}_{S_A}\,,\qquad \omega_A \equiv \frac{\partial m_A}{\partial S_A}{\bigg|}_{\mu_A}\,. 
\eeq
In particular, $\omega_A$ can be interpreted as the proper rotation frequency of body $A$. Assuming the validity of the Christodoulou mass formula $m_A^2 = \mu_A^2 + {S_A^2}/(4 \mu_A^2)$ for Kerr black holes \cite{Ch.70,ChRu.71}, \textit{i.e.} neglecting the influence of the other body, we have
\bseq
\begin{align}
c_A &= \frac{\mu_A}{m_A}\left(1-\frac{S_A^2}{4\mu_A^4}\right) , \label{cA} \\
\omega_A &= \frac{S_A}{4 m_A \mu_A^2} \,.
\end{align}
\eseq
For an isolated black hole, the coefficient $c_A$ is related to the constant surface gravity $\kappa_A$ by $c_A = 4 \mu_A \kappa_A$. Note that, at linear order in the spins, we have $c_A = 1 + \mathcal{O}(S_A^2)$. While the first law \eqref{1st_law} does not account for spin effects $\mathcal{O}(S_A^2)$ or higher, we shall not make the substitution $c_A \to 1$ in the equations and discussion below, in order to ease the comparison with the binary black hole case.

Identifying each spinning point mass with an ``extended body'' (or black hole) characterized by coefficients $c_A$ and $\omega_A$, we use the definitions \eqref{dmA} and find that the first law \eqref{1st_law} can be rewritten as 
\beq\label{1st_law_2PP}
	\delta M - \Omega \, \delta J = \sum_A \Bigl[ c_A z_A \, \delta \mu_A + \left( z_A \, \omega_A + \Omega_A - \Omega \right) \delta S_A \Bigr]\,. \\
\eeq
This variational relationship is reminiscent of the first law of binary black hole mechanics of Friedman, Ury{\=u}, and Shibata \cite{Fr.al.02}:
\beq
\label{1st_law_2BH}
\delta M - \Omega \, \delta J = \sum_A 4 \mu_A \kappa_A \, \delta \mu_A \, ,
\eeq
which holds for two actual black holes with irreducible masses $\mu_A$ and constant surface gravities $\kappa_A$ (the surface areas of the black holes being $\mathcal{A}_A=16\pi\mu_A^2$).

\subsection{First law for corotating binaries}
\label{subsec:1st_law}

Both first laws \eqref{1st_law_2PP} and \eqref{1st_law_2BH} have been derived for circular orbits and spins aligned with the orbital angular momentum. However, whereas Eq.~\eqref{1st_law_2PP} is valid for arbitrary spin magnitudes, Eq.~\eqref{1st_law_2BH} can only describe \textit{corotating} black holes. This key difference is intimately related to the fact that black holes are finite-sized objects, whereas point particles have (by definition) no spatial extension. The assumption of a circular orbit implies the existence of a helical symmetry. In the binary black hole case, the helical Killing field must be tangent to the null geodesic generators of the horizons, which entirely constrains the rotational state of each black hole. If corotation were not realized, the resulting non-vanishing shear would lead to the growth of the horizon's surface areas, in contradiction with the hypothesis of helical symmetry \cite{Fr.al.02}. In the binary point-particle case, however, no such restriction occurs because of the point-like nature of such idealized objects.

By analogy with Eq.~\eqref{1st_law_2BH}, we see from Eq.~\eqref{1st_law_2PP} that the two point particles will be ``corotating'' if and only if
\beq\label{corot}
	z_A \, \omega_A + \Omega_A = \Omega \,.
\eeq
This equation determines the values of the proper frequencies $\omega_A$ to be used to compute the corotating point particle spins $S_A$ through the relation $S_A = 4 m_A \mu_A^2 \omega_A$ (and the Christodoulou mass formula). Physically, the condition \eqref{corot} means that the redshifted proper rotation frequency of each particle, $z_A \,\omega_A$, must be equal to the circular-orbit frequency $\Omega$, as seen in a frame rotating at the angular rate $\Omega_A$ with respect to an inertial frame of reference. Alternatively, we can associate to the proper rotation frequency $\omega_A$ a proper rotation angle $\phi_A$ by the definition $\omega_A \equiv \ud \phi_A / \ud \tau_A$, such that Eq.~\eqref{corot} becomes $\ud \phi_A = \left( \Omega - \Omega_A \right) \ud t = \ud \varphi - \ud \Phi_A$. This relationship gives the change $\ud \phi_A$ in the proper rotation angle in terms of the changes $\ud \varphi$ and $\ud \Phi_A$ in the orbital phase and precession angle during a coordinate time interval $\ud t$.

For such corotating point particles, the first law \eqref{1st_law_2PP} simplifies considerably to
\beq\label{1st_law_corot}
	\delta M - \Omega \, \delta J = \sum_A c_A z_A \, \delta \mu_A \, .
\eeq
It is almost \textit{identical} to that derived for non-spinning binaries in Paper I, which is Eq.~\eqref{1st_law_corot} with the substitutions $\delta \mu_A \rightarrow \delta m_A$ and $c_A \rightarrow 1$. Since the irreducible mass $\mu_A$ of a rotating black hole is the spin-independent part of its total mass $m_A$, this observation suggests that corotating binaries are very similar to non-spinning binaries, at least from the perspective of the first law of mechanics. This helps to explain why the first law of binary black hole mechanics \eqref{1st_law_2BH} has been observed --- in the context of sequences of quasi-equilibrium initial data --- to be remarkably well satisfied by \textit{non}-spinning binary black holes \cite{Ca.al2.06}, even though it should, in principle, hold only for corotating black hole binaries \cite{Fr.al.02}.

We also note that the variational first law \eqref{1st_law_corot} admits the first integral
\beq\label{1st_integral_corot}
	M - 2 \Omega J = \sum_A \mu_A c_A z_A \, ,
\eeq
which again is very similar to the first integral relation derived in the non-spinning case, \textit{i.e.} Eq.~\eqref{1st_integral_corot} with $\mu_A \rightarrow m_A$ and $c_A \rightarrow 1$. Equation \eqref{1st_integral_corot} can also be derived from the first integral \eqref{1st_integral_ter} valid for arbitrary spins, in which we replace the masses $m_A$ by the irreducible masses $\mu_A$ and spins $S_A$ using $m_A = \mu_A c_A + 2 \omega_A S_A$ (the analogue of Smarr's formula \cite{Sm.73} for spinning point particles treated as Kerr black holes), and impose the condition \eqref{corot} for corotation.

Now, comparing the first law \eqref{1st_law_2BH} for corotating black holes with the first law \eqref{1st_law_corot} for corotating point particles, we notice the formal analogy
\beq\label{analogy}
	c_A z_A ~\longleftrightarrow~ 4 \mu_A \kappa_A \, .
\eeq
This point-particle/black-hole analogy was pointed out in Paper I (with $c_A \rightarrow 1$), in the case of non-spinning point masses. Up to the factors of $c_A = 1 + \mathcal{O}(S_A^2)$, the first law for corotating point particles expressed in terms of the irreducible masses $\mu_A$ is identical to the first law for non-spinning point particles expressed in terms of the total masses $m_A = \mu_A + \mathcal{O}(S_A^2)$. As derived in Paper I, the analogy \eqref{analogy}, with $c_A \rightarrow 1$, was not entirely physically motivated, because it required the identification of the irreducible masses of the corotating black holes with the total masses of the non-spinning point particles. Having been established here in the case of corotating point particles, and requiring only the identification of irreducible masses, the relation \eqref{analogy} is much more compelling.

In the limit of large separation, the black holes or point particles can be viewed, in first approximation, as isolated. In that limit, we know that $4 \mu_A \kappa_A \rightarrow c_A$ for each black hole, and $z_A \rightarrow 1$ for each point particle, such that the analogy \eqref{analogy} is consistent. Going beyond the large separation limit, Eq.~\eqref{analogy} suggests that the deviations of the redshifts of the point particles from one provide a measure of the interaction between the black holes. In particular, since the redshifts $z_A = \ud \tau_A / \ud t$ must be less than one, we expect
\beq\label{interaction}
\kappa_A < \frac{c_A}{4\mu_A}
\eeq
in binary systems of corotating black holes on circular orbits. In other words, the tidal interaction between the holes should \textit{decrease} the surface gravities with respect to their values in isolation. This prediction could be checked by computing numerically $\kappa_A$ in sequences of quasi-circular initial data relying on the existence of a global helical Killing vector \cite{Go.al.02,Gr.al.02}. One could also compare quantitatively PN results for $z_A$ \cite{De.08,Bl.al.10,Bl.al2.10,Le.al.12} with numerical results for $4 \mu_A \kappa_A$, as functions of the circular-orbit frequency $\Omega$.

\subsection{Proper rotation frequencies through 2PN order}
\label{subsec:omegaA}

The condition \eqref{corot} for corotation can be solved for the proper rotation frequencies $\omega_A$ of the particles:
\beq
\label{corot_bis}
\omega_A = u^t_A \left( \Omega - \Omega_A \right)\,,
\eeq
where $u^t_A = 1 / z_A$ has been computed up to very high PN orders, for \textit{non}-spinning binaries \cite{De.08,Bl.al.10,Bl.al2.10,Le.al.12}. Here we only need the 1PN-accurate results. For circular orbits, and in the center-of-mass frame, the spin-independent contributions to the redshift associated with particle $1$ read
\begin{align}\label{u1}
	u^t_1 &= 1 + \left( \frac{3}{4} + \frac{3}{4} \Delta - \frac{\nu}{2} \right) x\\
& + \left( \frac{27}{16} + \frac{27}{16} \Delta - \frac{5}{2} \nu - \frac{5}{8} \Delta \, \nu + \frac{\nu^2}{24} \right) x^2 + \mathcal{O}(x^3) \, .\nonumber
\end{align}
However, for corotating binaries we should also include spin contributions. The lowest order spin-related effect comes from the leading-order 1.5PN SO terms, as given by Eq.~\eqref{z1c}. Since $\chi_A \propto \Omega$ at leading order, this term will yield a 3PN contribution, which can be neglected here. Combining Eqs.~\eqref{Omega1} and \eqref{u1}, the condition \eqref{corot_bis} for corotation readily fixes at 2PN order
\beq
\label{omegaA}
\omega_A = \Omega \left\{ 1 - \nu \, x + \nu \left( - \frac{3}{2} + \frac{\nu}{3} \right) x^2 + \mathcal{O}(x^3) \right\} .
\eeq
Noticeably, even though $u^t_1 \neq u^t_2$ and $\Omega_1 \neq \Omega_2$, we find $\omega_1=\omega_2$ up to 2PN order included. Future work should investigate whether this symmetry property still holds at the next 3PN order. Note that in the Newtonian limit $x \to 0$ or the test-particle limit $\nu \to 0$ we simply have $\omega_A = \Omega$, in agreement with physical intuition. 

The observation that $\omega_1 = \omega_2$ up to high order suggests that the proper rotation frequency $\omega_A$ might physically correspond to the rotation rate of the tidal field of the companion $B \neq A$, as measured in the local asymptotic rest frame (LARF) of body $A$ \cite{ThHa.85}. By matching the near-zone PN metric of a binary system of point particles to the metric of two tidally distorted Schwarzschild black holes, Alvi \cite{Al.00} (see also Ref.~\cite{TaPo.08}) computed the 1PN-accurate expression of the rotation rate of the tidal field of body $B$ in the LARF of body $A$. Interestingly, his result agrees with the restriction at 1PN order of our 2PN-accurate expression \eqref{omegaA}.

Furthermore, Caudill \textit{et al.} \cite{Ca.al2.06} previously made that same hypothesis, in the context of quasi-equilibrium initial data for equal-mass black hole binaries. They found that using the 1PN-accurate result $\omega_A = \Omega \left( 1 - \nu \, x \right)$ instead of the leading-order result $\omega_A = \Omega$ considerably improved the agreement between two different quasi-local measures of the individual spins of the black holes; see Eqs.~(54)--(56) and Figs.~4 and 5 of Ref.~\cite{Ca.al2.06}. It would be interesting to revisit their work using our improved, 2PN-accurate expression \eqref{omegaA} for the proper rotation frequencies.

In Ref.~\cite{Bl.02}, the binding energy $E = M - m$ and total angular momentum $J$ of a binary system of corotating point particles were computed by replacing the masses $m_A$ and spins $S_A$ by the irreducible masses $\mu_A$ and proper rotation frequencies $\omega_A$, using the PN-expanded form of the relations between $(m_A,S_A)$ and $(\mu_A,\omega_A)$ obtained from the Christodoulou formula, and assuming the leading-order result $\omega_A = \Omega$. Using the corrected formula \eqref{omegaA} instead, the results of Ref.~\cite{Bl.02} are modified. Redoing the calculation for the \textit{additional} spin-related contributions $E^\text{cor}$ and $J^\text{cor}$ to the binding energy and total angular momentum in the corotating case, we find 
\bseq\label{EJcorot}
	\begin{align}
		E^\text{cor} &= \mu \left( 2 - 6 \eta \right) x_\mu^3 + \mu \eta \left( - 10 + 25 \eta \right) x_\mu^4 \, , \label{Mcorot} \\ 
		\Omega J^\text{cor} &= \mu \left( 4 - 12 \eta \right) x_\mu^3 + \mu \eta \left( - 16 + 40 \eta \right) x_\mu^4 \, , \label{Jcorot}
	\end{align}
\eseq
where the total mass $\mu = \mu_1 + \mu_2$, the symmetric mass ratio $\eta = \mu_1 \mu_2 / \mu^2$, and the dimensionless invariant PN parameter $x_\mu = (\mu\Omega)^{2/3}$ are now expressed in terms of the irreducible masses $\mu_A$, rather than the masses $m_A$. The 2PN and 3PN spin-related contributions \eqref{EJcorot} must be added to the known 3PN-accurate expressions of $E$ and $J$ for spinless binaries. As expected, the 1PN correction in Eq.~\eqref{omegaA} modifies the 3PN terms in \eqref{EJcorot} with respect to the results of Ref.~\cite{Bl.02}. It can be checked that the additional contributions \eqref{EJcorot} for corotating binaries now satisfy the relation
\beq\label{law}
	\frac{\partial E^\text{cor}}{\partial \Omega}\bigg|_{\mu_A} = \Omega \, \frac{\partial J^\text{cor}}{\partial \Omega}\bigg|_{\mu_A} .
\eeq
Since the spin-independent contributions to $E$ and $J$ are known to satisfy that same relation (see Paper I), we conclude that the thermodynamical law is verified, as it should be according to the first law \eqref{1st_law_corot}, for corotating point particles. It would be interesting to revisit the PN/NR comparison of Ref.~\cite{Bl.02} using the corrected formulas \eqref{EJcorot}.

\section{Summary and prospects}
\label{sec:concl}

In this paper we generalized to spinning point particles the first law of binary mechanics established in Paper I for non-spinning point masses. We derived a simple relation between the Hamiltonian and the redshift observable using a general formalism for the Fokker Lagrangian and Hamiltonian of a system of spinning particles. We then derived the first law within the canonical ADM Hamiltonian formalism, for binaries on circular orbits and with spins aligned or anti-aligned with the orbital angular momentum. We also obtained several useful relations linking the main quantities describing the two-body dynamics for circular orbits, mainly the ADM mass, angular momentum, and precession frequencies. 

Similarly to previous work \cite{Le.al2.12,Ba.al.12}, a calculation of the gravitational self-force effect on the redshift observable of a particle on a circular, equatorial orbit around a Kerr black hole \cite{Sh.al.12,Fr.al.12} could be used to compute the exact spin-orbit contributions to the binding energy and orbital angular momentum, for aligned or anti-aligned spins, beyond the test-particle approximation. This would allow computing the frequency shift of the Kerr innermost stable circular orbit under the effect of the conservative self-force (at linear order in the spin), an important strong-field benchmark.

New comparisons to numerical relativity simulations of spinning black hole binaries could then explore further the promise of using perturbation theory to model comparable-mass compact binaries \cite{Le.al.11,Le.al2.12}. This information could also be used to improve the EOB model for spinning binaries \cite{Da.01,Da.al2.08,BaBu.10,Na.11,BaBu.11,Ta.al.12}. Furthermore, once second-order gravitational self-force calculations \cite{De.12,Po.12,Gr.12,Po2.12} become mature enough, these same ideas could be applied to compute the fully relativistic second-order contributions in the (symmetric) mass ratio to the binding energy and angular momentum, for circular orbits. All of these results would be valuable to improve template waveforms for inspiraling and coalescing binaries of compact objects.

Moreover, our simple relation between the redshift observable and the (ADM, or more generally Fokker-type) Hamiltonian allowed us to compute spin-orbit effects in the redshift through 2.5PN order, for circular orbits and spins aligned or anti-aligned. These results agree with a recent direct calculation based on the PN metric \cite{Fr.al.12}. Future work can extend the knowledge of the redshift function to higher PN order using the recently computed spin-orbit effects at 3.5PN order \cite{HaSt.11,Ma.al.12}. It would also be interesting and useful to extend the first law to second order in the spins, in order to account for spin-spin contributions and quadrupolar deformations.

Finally, we specified our first law for binary systems of spinning point particles to the corotating case, allowing a comparison with the binary black hole case. Extending previous results \cite{Al.00,TaPo.08}, we computed the proper rotation frequencies of the particles through 2PN order. This finding could be used to improve the quasi-local measure of the individual spins of black holes in the context of quasi-equilibrium initial data \cite{Ca.al2.06}.

\acknowledgments
It is a pleasure to thank E. Barausse and J.~Steinhoff for useful discussions. A.B. acknowledges partial support from the NSF Grants No. PHY-0903631 and PHY-1208881, and from the NASA Grant NNX09AI81G. A.L.T. acknowledges support from the NSF Grants No. PHY-0903631 and PHY-1208881, and from the Maryland Center for Fundamental Physics. A.B. and A.L.T. also thank the Kavli Institute for Theoretical Physics (supported by the NSF Grant No. PHY11-25915) for hospitality during the preparation of this manuscript.

\bibliography{}

\end{document}